\documentclass[referee,usenatbib]{mnras}
\usepackage[T1]{fontenc}

\DeclareRobustCommand{\VAN}[3]{#2}
\let\VANthebibliography\thebibliography
\def\thebibliography{\DeclareRobustCommand{\VAN}[3]{##3}\VANthebibliography}

\usepackage{graphicx}	
\usepackage{amsmath}	
\usepackage{amssymb}	
\usepackage{lastpage}   

\title[Chandra Spectroscopy of Supernova Remnant N63A]{A Chandra X-ray Study of Supernova Remnant N63A in the Large Magellanic Cloud}

\author[Karag\"oz et al.]{
E. Karag\"oz,$^{1}$\thanks{E-mail: emrekaragoz1@ogr.iu.edu.tr}
N. Alan,$^{2}$
S. Bilir,$^{2}$
and
S. Ak$^{2}$
\\
$^1$Istanbul University, Institute of Graduate Studies in Science, Programme of Astronomy and Space Sciences, 34116, \\
Beyaz{\i}t, Istanbul, Turkey\\
$^2$Istanbul University, Faculty of Science, Department of Astronomy and Space
Sciences, 34119, Beyaz{\i}t, Istanbul, Turkey\\
}
\date{Accepted XXX. Received YYY; in original form ZZZ}

\pubyear{2015}

\begin{document}
\label{firstpage}
\pagerange{\pageref{firstpage}--\pageref{lastpage}}
\maketitle

\begin{abstract}

We perform extensive spectroscopy of the supernova remnant N63A in the Large Magellanic Cloud, using $\sim 43$ ks {\it Chandra} archival data. By analysing the spectra of the entire remnant, we determine the abundance distributions for O, Ne, Mg, Si, and Fe. We detect evidence of enhanced O and possibly Ne and Mg in some of the central regions which might indicate an asymmetric distribution of the ejecta. The average O/Ne, O/Mg, and Ne/Mg abundance ratios of the ejecta are in plausible agreement with the nucleosynthesis products from the explosion of a $\sim40$ $M_{\odot}$ progenitor. We estimate an upper limit on the Sedov age of $\sim 5,400\pm200$ yr and explosion energy of $\sim 8.9\pm 1.6\times 10^{51}$ erg for N63A. We discuss the implications of our results for the morphological structure of the remnant, its circumstellar medium and the nature of the progenitor star.

\end{abstract}




\section{Introduction}

Supernova (SN) explosions, the most energetic stellar events known, play a crucial role in shaping the energy density, chemical enrichment, and evolution of galaxies. Core-collapse explosions of massive stars ($M>8 M_{\odot}$) account for $\sim 3/4$ of all supernovae (SNe) \citep{Tsujimoto95, Sato07}, and their remnants precious tools for understanding the recent star formation, the dynamics of supernova explosions, the composition of the interstellar medium (ISM), and nature of the progenitor. The structure of supernova remnants (SNRs) and their interactions with the ambient medium provides insight into their origin and effects on their environment. N63A is one of the brightest SNRs in the Large Magellanic Cloud \citep[LMC;][]{Westerlund66}, and provides an excellent laboratory for investigating such structures and interactions.

N63A, first identified as an SNR by \citet{Mathewson64}, appears to be embedded within the H II region N63 coincident with the OB association NGC 2030 or LH83 \citep{Lucke70, Kennicutt88}. For this reason, it is believed that the remnant to be the product of the SN explosion of one of the most massive Population I stars in the dense and complex NGC 2030 or LH83 O-B association \citep{Lucke70, vandenBergh80, Shull83, Chu88, Hughes98}. The core-collapse origin of N63A was also corroborated by detailed measurements of Fe K$\alpha$ centroids \citep{Yamaguchi14}. \citet{Oey96} designated that the currently most luminous star in NGC 2030 is an O7 star with a mass of $\sim40 M_{\odot}$, therefore the progenitor of N63A’s supernova was probably more massive than this and with a main sequence spectral type earlier than O7. Furthermore, observational studies have confirmed that N63A is the first SNR formed in the HII region \citep{Shull83}. The age of the SNR is estimated to be $\sim 4,500$ yr \citep*{Williams06} and 2,000-5,000 yr \citep*{Hughes98, Warren03}, indicating that the natal gas may still be associated with N63A.

The size of N63A in X-rays is $r\sim 81'' \times 67''$ or $\sim 18$ pc in diameter, assuming a distance of 50 kpc \citep[e.g.][]{Dickel93, Feast99} and this size is approximately three times that of the optical remnant containing three prominent lobes \citep{Mathewson83}. The two eastern lobes with high-intensity ratios of [S II]/H$\alpha$ \citep{Payne08} represent the shock-heated gas, while the third western lobe with a low-intensity ratio corresponds to the photoionized H II region \citep{Levenson95}. All optical lobes show molecular shock features with their near-infrared colours, suggesting that the shocked molecular gas dominates in the remnant \citep{Williams06}. Subsequent detailed infrared spectroscopy verified that shock-excited molecular hydrogen lines are determined in all-optical lobes \citep{Caulet12}. Similar to the fact that the X-ray emission from N63A is also consistent with swept-up ISM \citep{Hughes98}, the shock heating is the result of interactions with the ISM, rather than with SN ejecta, based on the derived abundances \citep{Russell90}. The imaging spectroscopy of X-rays also indicates the presence of dense interstellar gas with a mass of at least $\sim450M_{\odot}$ \citep{Warren03}. \citet{Sano19} using the ALMA data calculated the total mass of the molecular clouds is $\sim800 M_{\odot}$ for the shock-ionized region and $\sim1,700$ $M_{\odot}$ for the photoionized region. They also reveal that the absorbing column densities toward the molecular clouds are $\sim 1.5-6.0\times 10^{21}$ cm$^{-2}$.

In previous X-ray studies, several sub-regions characteristically representing emission from the swept-up ambient gas and the shocked ejecta were examined, but an extensive X-ray study of the entire remnant has not yet been performed. In this work, based on the archival {\it Chandra} data, we perform the spatially-resolved spectral analysis of the entire SNR to provide unprecedented details on radial and azimuthal structures of N63A. We describe the X-ray data and the data reduction in Section 2. We present the X-ray imaging and spectral analysis of the SNR in Section 3. We discuss the results and compare this work to other abundance measurements in Section 4.

\section{X-ray data and reduction}
We used archival {\it Chandra} data (43.35 ks) of SNR N63A obtained on 2000 October 16 with the S3 chip Advanced Charged Couple Device Imaging Spectrometer (ACIS-S) detector array \citep{Bautz98}. We reprocessed observational data with CIAO version 4.13 via the \texttt{chandra\_repro} script. We removed time intervals that shows high particle background fluxes (by a factor of $\sim2$ higher than the mean background). After the data reduction, the total effective exposure time is $\sim 41$ ks.

\section{Analysis and Results}
\subsection{Imaging}
We created an ACIS-S3 broadband and an X-ray three-colour (RGB) images of N63A using archival {\it Chandra} data as shown in Figure 1. The X-ray morphology of the N63A shows a bright, sharp outer rim in the northwest, northeast, and southeast while a faint, diffuse in the southwest. There are also smaller diffuse ``crescent-like'' structures on the northern and eastern borders of the remnant. One of the remarkable structures on the N63A is the faint triangular ``hole-like'' area in just west of the geometrical centre of the remnant (Figure 1a). This area corresponds to optical lobes which are also coincident with the brightest emission in radio.

In Figure 1b, we show the {\it Chandra} RGB image with red, green and blue corresponding to the soft (0.30-1.11 keV), medium  (1.11-2.10 keV), and hard (2.10-7.00 keV) energy bands, respectively. The energy bands displayed in each colour were chosen to emphasise atomic line emission that illustrates the distribution of electron temperatures and ionization states across N63A. The major emission lines O and Ne are grouped together. We used the sub-band images with the native pixel scale of the ACIS detector ($0''.49$ pixel$^{-1}$) and then adaptively smoothed them. Although soft X-ray (red) dominates N63A, inner parts of the remnant generally show harder X-ray emission (blue). The RGB image showed that both the crescent-like regions and the entire rim of the remnant have softer spectra than average for N63A. There are also small blue clumps on the remnant that show a harder-than-average X-ray emission. 

\begin{figure*}
\begin{center}
\includegraphics[width=.96\textwidth]{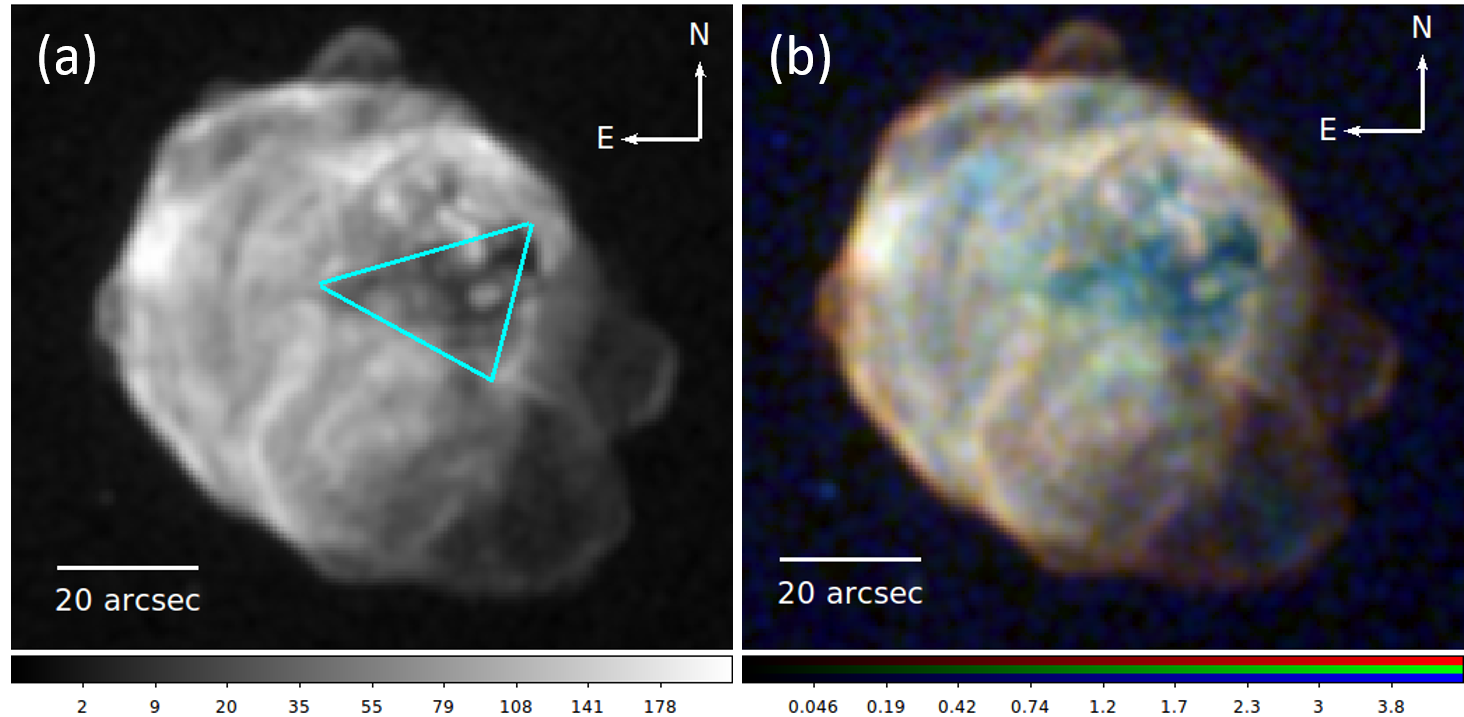}
\caption{(a) The broadband ACIS image (0.30-7.0 keV) and (b) the three-colour image of N63A: Red = 0.30-1.11 keV, green = 1.11-2.10 keV, and blue = 2.10-7.00 keV. The cyan triangle in the left panel represents the triangular hole-like structure. For the purposes of display, both images have been smoothed by a Gaussian kernel of $\sigma=0''.75$.}
\end{center}
\end{figure*}

\begin{figure}
\begin{center}
\includegraphics[scale=0.75, angle=0]{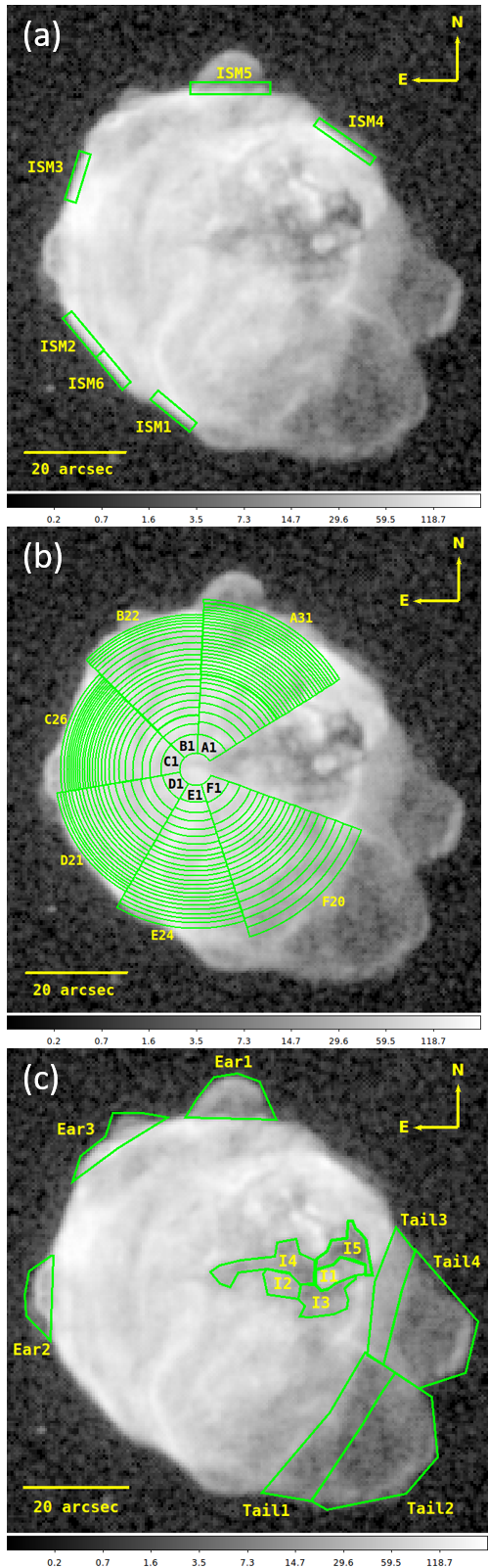}\\
\caption{Logarithmic-scale broadband image of N63A in the $0.3-7.0$ keV photon energy band. (a) The outermost six shell regions that are used to characterise the spectral nature of the swept-up ISM are marked. (b) The radial and azimuthal regions used for the spectral analysis are marked. (c) The inner and outer faint regions used for the spectral analysis are marked. All images have been smoothed by a Gaussian kernel of $\sigma=0''.25$.} 
\end{center}
\end{figure}

\subsection{Spectroscopy}

We first analysed the outer rim of SNR N63A, which represents the ISM swept up by forward shock. For this purpose, we selected six regions and marked them as ISM1-ISM6 (Figure 2a). Shell regions contain almost 5,000 counts on average, in the 0.3-3.0 keV energy band. In order to reveal the spatial structure of X-ray emission from metal-rich ejecta gas, we selected 144 regions through seven radial directions across N63A. We also selected 12 X-ray faint regions from the inner and outer parts of the remnant. The radial regions and directions are marked as shown in Figure 2b. Each region contains about 5,000 counts in the 0.3-3.0 keV energy band. The outer and inner faint X-ray regions that contain almost 4,000-6,000 counts are also marked as shown in Figure 2c. Then, we extracted the spectra from the observational data for each selected faint region using the CIAO script \texttt{specextract}. We performed the background subtraction using the spectrum extracted from source-free regions outside of N63A. We binned each extracted spectrum to comprise at least 20 counts per photon energy channel. We fit each regional spectrum with a non-equilibrium ionization (NEI) plane-shock model \citep[\texttt{vpshock} in XSPEC;][]{Borkowski01} with two foreground absorption column components, one for the Galactic ($N_{{\rm H, Gal}}$) and the other for the LMC ($N_{{\rm H, LMC}}$). We used NEI version 3.0.4 in XSPEC related to ATOMDB \citep{Foster12}, which was augmented to include inner shell lines, and updated Fe-L lines \citep[see, ][]{Badenes06}. We fixed $N_{{\rm H,Gal}}$ at $1.72\times10^{21}$ cm$^{-2}$ for the direction toward N63A \citep{HI4PI16} with solar abundances \citep{Anders89}. We fitted $N_{{\rm H,LMC}}$ assuming the LMC abundances \citep{Russell92, Schenck16}. We also fixed the redshift parameter at $z=8.75\times10^{-4}$ for the radial velocity (262.2 kms$^{-1}$) of the LMC \citep{McConnachie12}.

\subsubsection{Ambient Medium}

We fit ISM1 - ISM6 regions spectra using a one-component plane-parallel shock (\texttt{phabs $\times$ vphabs $\times$ vpshock}) model. In Figure 3, we show the spectra of ISM1 and ISM3 regions with models and residuals as a sample. We initially fixed all elemental abundances at the LMC values, i.e. He = 0.89, C=0.303, S=0.31, N=0.123, Ar=0.537, Ca=0.339, Ni=0.618 \citep{Russell92}, and O=0.13, Ne=0.20, Mg=0.20, Si=0.28, Fe=0.15 \citep{Schenck16}, in the plane-shock model. Hereafter, abundances are with respect to solar values \citep{Anders89}. We varied electron temperature ($kT$, where $k$ is the Boltzmann constant), ionization timescale ($n_{\rm e}t$, where $n_{\rm e}$ is the post-shock electron density, and $t$ is the time since the gas has been shocked) and the $N_{\rm H, LMC}$ absorbing column in the LMC. The normalisation parameter (a scaled volume emission measure, $EM=n_{\rm e}n_{\rm H}V$, where $n_{\rm H}$ is the postshock H density, and $V$ is the emission volume) is also varied. The reduced chi-square ($\chi^2_{\nu}$) values between $1.3-1.9$ for the model fits. Then to improve the fits we varied O, Ne, Mg, Si, Fe abundances and obtained the best-fit models for the shell regions ($\chi^2_{\nu}=1.01-1.50$). While the fitted Ne, Mg, and Si abundances are consistent (within statistical uncertainties) with values given by \citet{Russell92}, the fitted abundances for O and Fe are lower by a factor of $\sim  2-3$ than \citet{Russell92} values. Besides this all fitted elemental abundances are consistent with \citet{Schenck16} values within statistical uncertainties. We found that the outer rim spectrum of N63A is dominated by emission from the shocked low-abundant LMC ISM rather than that from the shocked metal-rich ejecta gas. The best-fit model parameters of the shell regions and their median values are listed in Table \ref{tab:shell_regions}.

\begin{figure*}
\begin{center}
\includegraphics[width=\textwidth]{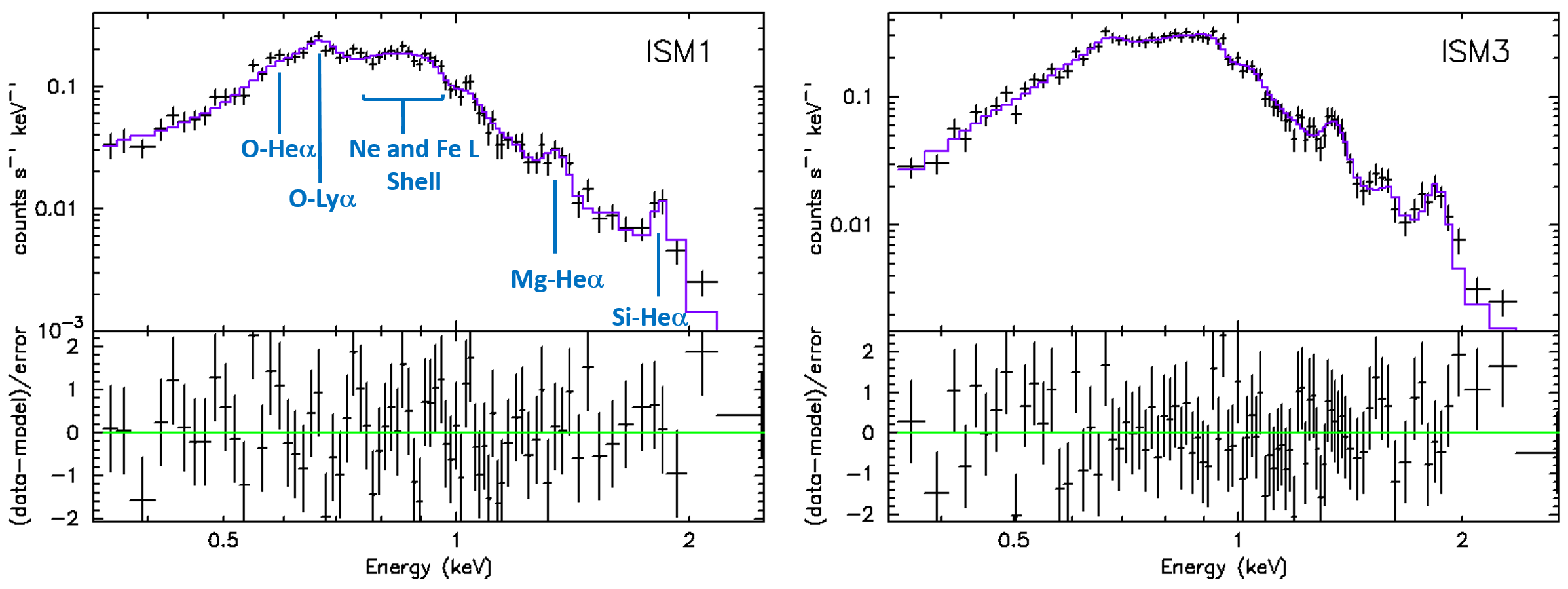}
\caption{Best-fit spectral models and residuals of the X-ray spectra from the ISM1 and ISM3 regions. Several atomic emission line features are marked on left panel.} 
\end{center}
\end{figure*}

\begin{table*}
\setlength{\tabcolsep}{5pt}
\renewcommand{\arraystretch}{1.5}
\scriptsize
  \centering
  \caption{Summary of spectral model fits to the six shell regions of N63A. The median shell values are given in the last line.}
    \begin{tabular}{ccccccccccc}
\hline
Region    &    $n_{\rm H}$             &   $kT$  & $n_{\rm e}t$                & $EM$                 & O    & Ne  & Mg  & Si & Fe & $\chi^{2}_{\nu}$\\
    &($10^{22}$cm$^{-2}$)  &  (keV)  & ($10^{11}$cm$^{-3}$s) & ($10^{57}$cm$^{-3}$) &      &     &     &    &    &                 \\
\hline
ISM1 & $0.04_{-0.03}^{+0.08}$ & $0.61_{-0.14}^{+0.11}$ & $ 1.06_{-0.46}^{+1.96}$ & $ 13.07_{-3.32}^{+20.81}$ & $0.16_{-0.03}^{+0.03}$ & $0.28_{-0.05}^{+0.06}$ & $0.21_{-0.08}^{+0.10}$ & $0.40_{-0.14}^{+0.22}$ & $0.15_{-0.05}^{+0.05}$ & 1.25 \\
ISM2 & $0.01_{-0.01}^{+0.01}$ & $0.60_{-0.10}^{+0.06}$ & $ 1.53_{-0.40}^{+1.80}$ & $ 18.95_{-2.75}^{+27.11}$ & $0.16_{-0.03}^{+0.04}$ & $0.25_{-0.05}^{+0.05}$ & $0.21_{-0.08}^{+0.06}$ & $0.28_{-0.12}^{+0.12}$ & $0.15_{-0.04}^{+0.02}$ & 1.40 \\
ISM3 & $0.01_{-0.01}^{+0.07}$ & $0.57_{-0.07}^{+0.05}$ & $ 2.88_{-0.98}^{+1.70}$ & $ 24.82_{-4.19}^{+35.34}$ & $0.17_{-0.05}^{+0.04}$ & $0.36_{-0.08}^{+0.06}$ & $0.29_{-0.08}^{+0.08}$ & $0.38_{-0.11}^{+0.13}$ & $0.13_{-0.03}^{+0.02}$ & 1.01 \\
ISM4 & $0.01_{-0.01}^{+0.02}$ & $0.50_{-0.08}^{+0.13}$ & $ 1.77_{-1.03}^{+3.15}$ & $ 17.94_{-6.39}^{+26.15}$ & $0.16_{-0.02}^{+0.04}$ & $0.30_{-0.05}^{+0.06}$ & $0.28_{-0.09}^{+0.10}$ & $0.26_{-0.14}^{+0.18}$ & $0.08_{-0.02}^{+0.04}$ & 1.22 \\
ISM5 & $0.01_{-0.02}^{+0.04}$ & $0.58_{-0.10}^{+0.16}$ & $ 1.49_{-0.84}^{+0.76}$ & $ 13.17_{-4.48}^{+19.16}$ & $0.20_{-0.05}^{+0.02}$ & $0.30_{-0.06}^{+0.07}$ & $0.25_{-0.08}^{+0.08}$ & $0.42_{-0.16}^{+0.15}$ & $0.16_{-0.04}^{+0.07}$ & 1.50 \\
ISM6 & $0.01_{-0.02}^{+0.04}$ & $0.59_{-0.10}^{+0.13}$ & $ 1.16_{-0.55}^{+1.25}$ & $ 15.65_{-4.46}^{+21.86}$ & $0.14_{-0.02}^{+0.02}$ & $0.28_{-0.05}^{+0.06}$ & $0.20_{-0.07}^{+0.08}$ & $0.40_{-0.15}^{+0.13}$ & $0.10_{-0.03}^{+0.04}$ & 1.50 \\
\hline
Median & $0.01_{-0.02}^{+0.04}$ & $0.59_{-0.10}^{+0.12}$ & $ 1.51_{-0.70}^{+1.75}$ & $16.80_{-4.33}^{+24.01}$ & $0.16_{-0.03}^{+0.04}$ & $0.29_{-0.05}^{+0.06}$ & $0.23_{-0.08}^{+0.08}$ & $0.39_{-0.14}^{+0.14}$ & $0.14_{-0.04}^{+0.04}$ & ---  \\
\hline
    \end{tabular}%
      \label{tab:shell_regions}%
\\
Note: Abundances are with respect to solar \citep{Anders89}. Uncertainties are at the 90\% confidence level. The Galactic column $N_{{\rm H,Gal}}$ is fixed at $1.72\times10^{21}$ cm$^{-2}$ \citep{HI4PI16}. 
\end{table*}%

\subsubsection{Metal-Rich Ejecta}
We perform an extensive spatially resolved spectral analysis of its X-ray emission based on 144 radial and azimuthal regional (see Figure 2b) spectra to study the detailed spatial distribution of metal-rich ejecta in N63A. The outer regions of the remnant can be modelled with a single-component plane shock model, and the spectral parameters for these regions are compatible with median shell values. The spectra of central parts of the remnant cannot be fit by a single shock model with abundances fixed at the values that we estimated for the mean shell (i.e. $\chi^2_{\nu}>2.0$). These show the existence of an additional shock component, likely representing the emission from the shocked metal-rich ejecta gas, superposed with the projected shell emission. Therefore, we performed a two-component NEI shock model (\texttt{phabs $\times$ vphabs $\times$ (vpshock+vpshock)}) to fit these spectra, one for the underlying mean shell spectrum and the other responsible for the metal-rich ejecta component. We fixed $N_{{\rm H, LMC}}$ at the mean shell value. We also fixed all model parameters, except for normalisation, of the underlying swept-up ISM component at the values for the mean shell (Table \ref{tab:shell_regions}). For the second shock component, we first varied $kT$, $n_{\rm e}t$, and normalisation and fixed the elemental abundances at the mean shell values. The fit for each regional spectrum was not statistically acceptable ($\chi^2_{\nu}>2.0$) because the model was not able to reproduce emission lines from various elements. Then we thawed elemental abundances for the second component to improve each regional spectral fit. With abundances for O, Ne, Mg, Si, and Fe varied, our spectral model fits significantly improved ($\chi^2_{\nu}<1.6$). In Figures 4-5, we show some example spectra extracted from the regions marked in Figure 2b with best-fit models and residuals. The O, Ne and Mg in some of the central parts of the remnant in directions A and F are enhanced compared to those of the mean shell values. Si and Fe abundances are generally consistent with mean shell values within statistical uncertainties. The best-fit model parameters for radial regions are listed in Table \ref{tab:six_directions}, and radial profiles of spectral parameters are shown in Figures 7-9.

\begin{table*}
\setlength{\tabcolsep}{3.5pt}
\renewcommand{\arraystretch}{1.4}
\scriptsize
  \centering
  \caption{Spectral parameters from NEI shock model fits for six radials and different number of azimuthal regions spectra of N63A.}
    \begin{tabular}{cccccccccccc}
\hline
Region &   $r$    & $n_{\rm H}$            & $kT$                   & $n_{\rm e}t$             & $EM$                   & O                      & Ne                     & Mg                     & Si                     & Fe         & $\chi^{2}_{\nu}$\\
       &  ($''$)  & ($10^{22}$cm$^{-2}$)   & (keV)                  & ($10^{11}$cm$^{-3}$s)    & ($10^{57}$cm$^{-3}$)   &                        &                        &                        &                        &                        & \\
\hline
A01 (*)	&	3.46	&	$0.02_{-0.02}^{+0.04}$	&	$1.10_{-0.20}^{+2.53}$	&	$5.92_{-2.47}^{+6.82}$	&	$17.06_{-1.78}^{+1.60}$	&	$0.44_{-0.44}^{+0.25}$	&	$0.51_{-0.47}^{+4.23}$	&	$0.64_{-0.28}^{+4.36}$	&	$0.27_{-0.17}^{+0.27}$	&	$0.23_{-0.23}^{+0.18}$	&	1.00	\\
A02 (*)	&	6.56	&	$0.01_{-0.01}^{+0.03}$	&	$1.14_{-0.08}^{+0.10}$	&	$5.23_{-1.69}^{+3.47}$	&	$14.52_{-1.44}^{+1.37}$	&	$0.82_{-0.26}^{+0.36}$	&	$0.96_{-0.64}^{+0.98}$	&	$0.80_{-0.32}^{+0.56}$	&	$0.34_{-0.19}^{+0.30}$	&	$0.38_{-0.16}^{+0.17}$	&	1.34	\\
A03 (*)	&	8.81	&	$0.01_{-0.01}^{+0.02}$	&	$0.81_{-0.05}^{+0.06}$	&	$8.18_{-1.75}^{+2.92}$	&	$18.15_{-3.26}^{+3.51}$	&	$0.61_{-0.15}^{+0.15}$	&	$0.59_{-0.18}^{+0.22}$	&	$0.35_{-0.14}^{+0.17}$	&	$0.33_{-0.11}^{+0.15}$	&	$0.12_{-0.03}^{+0.03}$	&	1.16	\\
A04 (*)	&	10.69	&	$0.01_{-0.01}^{+0.02}$	&	$0.81_{-0.17}^{+0.04}$	&	$9.28_{-2.44}^{+4.62}$	&	$20.00_{-3.27}^{+3.35}$	&	$0.43_{-0.18}^{+0.08}$	&	$0.82_{-0.39}^{+0.11}$	&	$0.52_{-0.29}^{+0.03}$	&	$0.53_{-0.24}^{+0.10}$	&	$0.16_{-0.08}^{+0.02}$	&	1.00	\\
A05 (*)	&	12.34	&	$0.01_{-0.01}^{+0.01}$	&	$0.83_{-0.07}^{+0.10}$	&	$12.83_{-4.95}^{+19.87}$	&	$19.23_{-2.64}^{+2.41}$	&	$0.44_{-0.16}^{+0.19}$	&	$0.41_{-0.28}^{+0.25}$	&	$0.47_{-0.17}^{+0.29}$	&	$0.30_{-0.13}^{+0.18}$	&	$0.17_{-0.05}^{+0.09}$	&	1.06	\\
A06 (*)	&	13.76	&	$0.02_{-0.02}^{+0.03}$	&	$0.75_{-0.05}^{+0.05}$	&	$8.33_{-0.88}^{+6.83}$	&	$22.43_{-2.66}^{+3.86}$	&	$0.37_{-0.08}^{+0.10}$	&	$0.31_{-0.12}^{+0.13}$	&	$0.35_{-0.13}^{+0.15}$	&	$0.32_{-0.09}^{+0.10}$	&	$0.13_{-0.02}^{+0.03}$	&	1.22	\\
A07	&	14.96	&	$0.01_{-0.01}^{+0.01}$	&	$0.68_{-0.05}^{+0.06}$	&	$7.82_{-3.35}^{+5.39}$	&	$23.93_{-3.41}^{+4.85}$	&	$0.28_{-0.09}^{+0.08}$	&	$0.34_{-0.11}^{+0.13}$	&	$0.20_{-0.07}^{+0.08}$	&	$0.20_{-0.08}^{+0.09}$	&	$0.12_{-0.02}^{+0.03}$	&	1.59	\\
A08	&	15.96	&	$0.01_{-0.01}^{+0.01}$	&	$0.72_{-0.02}^{+0.08}$	&	$4.57_{-2.00}^{+3.52}$	&	$22.91_{-3.66}^{+3.58}$	&	$0.20_{-0.06}^{+0.11}$	&	$0.21_{-0.07}^{+0.08}$	&	$0.12_{-0.05}^{+0.05}$	&	$0.16_{-0.07}^{+0.07}$	&	$0.13_{-0.02}^{+0.03}$	&	1.59	\\
A09	&	16.79	&	$0.01_{-0.01}^{+0.01}$	&	$0.66_{-0.06}^{+0.06}$	&	$7.42_{-3.21}^{+5.28}$	&	$24.61_{-3.52}^{+3.93}$	&	$0.22_{-0.07}^{+0.10}$	&	$0.25_{-0.09}^{+0.09}$	&	$0.20_{-0.07}^{+0.07}$	&	$0.27_{-0.08}^{+0.09}$	&	$0.13_{-0.02}^{+0.03}$	&	1.38	\\
A10	&	17.51	&	$0.01_{-0.01}^{+0.01}$	&	$0.69_{-0.03}^{+0.03}$	&	$7.30_{-2.41}^{+5.81}$	&	$22.65_{-1.49}^{+3.73}$	&	$0.27_{-0.08}^{+0.13}$	&	$0.23_{-0.09}^{+0.12}$	&	$0.16_{-0.06}^{+0.07}$	&	$0.24_{-0.07}^{+0.08}$	&	$0.16_{-0.03}^{+0.03}$	&	1.55	\\
A11	&	18.21	&	$0.01_{-0.01}^{+0.02}$	&	$0.71_{-0.04}^{+0.07}$	&	$5.96_{-2.89}^{+3.37}$	&	$24.55_{-3.73}^{+1.76}$	&	$0.22_{-0.08}^{+0.09}$	&	$0.15_{-0.08}^{+0.05}$	&	$0.20_{-0.06}^{+0.06}$	&	$0.18_{-0.07}^{+0.04}$	&	$0.15_{-0.02}^{+0.04}$	&	0.98	\\
A12	&	18.91	&	$0.01_{-0.01}^{+0.01}$	&	$0.66_{-0.04}^{+0.06}$	&	$7.32_{-3.30}^{+6.02}$	&	$25.52_{-3.83}^{+3.54}$	&	$0.22_{-0.07}^{+0.11}$	&	$0.23_{-0.08}^{+0.09}$	&	$0.15_{-0.06}^{+0.07}$	&	$0.23_{-0.08}^{+0.09}$	&	$0.14_{-0.02}^{+0.03}$	&	1.26	\\
A13	&	19.64	&	$0.01_{-0.01}^{+0.02}$	&	$0.68_{-0.02}^{+0.04}$	&	$17.22_{-8.23}^{+11.02}$	&	$22.45_{-3.36}^{+2.74}$	&	$0.50_{-0.16}^{+0.23}$	&	$0.34_{-0.13}^{+0.13}$	&	$0.24_{-0.08}^{+0.10}$	&	$0.27_{-0.08}^{+0.10}$	&	$0.17_{-0.02}^{+0.04}$	&	1.25	\\
A14	&	20.39	&	$0.01_{-0.01}^{+0.01}$	&	$0.68_{-0.03}^{+0.05}$	&	$7.87_{-3.39}^{+4.49}$	&	$20.73_{-3.01}^{+1.87}$	&	$0.28_{-0.09}^{+0.12}$	&	$0.29_{-0.10}^{+0.12}$	&	$0.28_{-0.07}^{+0.10}$	&	$0.32_{-0.10}^{+0.11}$	&	$0.18_{-0.03}^{+0.04}$	&	1.08	\\
A15	&	21.09	&	$0.01_{-0.01}^{+0.01}$	&	$0.70_{-0.06}^{+0.06}$	&	$6.59_{-3.03}^{+5.85}$	&	$21.46_{-3.02}^{+3.33}$	&	$0.21_{-0.07}^{+0.12}$	&	$0.19_{-0.09}^{+0.11}$	&	$0.18_{-0.07}^{+0.07}$	&	$0.26_{-0.09}^{+0.10}$	&	$0.16_{-0.03}^{+0.04}$	&	1.25	\\
A16	&	21.76	&	$0.01_{-0.01}^{+0.04}$	&	$0.66_{-0.04}^{+0.04}$	&	$9.83_{-4.71}^{+4.94}$	&	$21.58_{-2.72}^{+4.45}$	&	$0.29_{-0.12}^{+0.12}$	&	$0.31_{-0.12}^{+0.11}$	&	$0.27_{-0.09}^{+0.09}$	&	$0.33_{-0.10}^{+0.11}$	&	$0.19_{-0.03}^{+0.03}$	&	1.20	\\
A17	&	22.44	&	$0.01_{-0.01}^{+0.01}$	&	$0.70_{-0.04}^{+0.05}$	&	$10.53_{-4.87}^{+9.35}$	&	$20.03_{-2.69}^{+2.97}$	&	$0.42_{-0.15}^{+0.24}$	&	$0.25_{-0.12}^{+0.15}$	&	$0.29_{-0.09}^{+0.11}$	&	$0.27_{-0.09}^{+0.10}$	&	$0.17_{-0.03}^{+0.04}$	&	1.25	\\
A18	&	23.10	&	$0.01_{-0.01}^{+0.04}$	&	$0.64_{-0.04}^{+0.04}$	&	$9.84_{-3.96}^{+5.45}$	&	$21.56_{-3.39}^{+4.34}$	&	$0.32_{-0.11}^{+0.15}$	&	$0.37_{-0.12}^{+0.18}$	&	$0.29_{-0.09}^{+0.10}$	&	$0.39_{-0.10}^{+0.14}$	&	$0.17_{-0.03}^{+0.04}$	&	1.36	\\
A19	&	23.79	&	$0.01_{-0.01}^{+0.05}$	&	$0.76_{-0.05}^{+0.08}$	&	$4.94_{-2.65}^{+4.02}$	&	$16.77_{-1.67}^{+3.63}$	&	$0.37_{-0.16}^{+0.22}$	&	$0.19_{-0.12}^{+0.13}$	&	$0.24_{-0.09}^{+0.11}$	&	$0.29_{-0.10}^{+0.12}$	&	$0.19_{-0.04}^{+0.05}$	&	0.91	\\
A20	&	24.53	&	$0.01_{-0.01}^{+0.01}$	&	$0.65_{-0.05}^{+0.06}$	&	$6.82_{-2.82}^{+4.98}$	&	$24.08_{-3.45}^{+3.78}$	&	$0.28_{-0.08}^{+0.12}$	&	$0.26_{-0.08}^{+0.10}$	&	$0.15_{-0.06}^{+0.07}$	&	$0.30_{-0.09}^{+0.11}$	&	$0.13_{-0.02}^{+0.03}$	&	1.22	\\
A21	&	25.26	&	$0.01_{-0.01}^{+0.01}$	&	$0.72_{-0.05}^{+0.06}$	&	$4.93_{-2.03}^{+3.56}$	&	$20.39_{-2.94}^{+3.24}$	&	$0.28_{-0.08}^{+0.13}$	&	$0.21_{-0.08}^{+0.09}$	&	$0.17_{-0.06}^{+0.07}$	&	$0.20_{-0.08}^{+0.09}$	&	$0.14_{-0.03}^{+0.03}$	&	1.05	\\
A22	&	25.99	&	$0.01_{-0.01}^{+0.01}$	&	$0.72_{-0.05}^{+0.06}$	&	$6.82_{-2.97}^{+5.05}$	&	$21.18_{-2.86}^{+2.22}$	&	$0.38_{-0.12}^{+0.18}$	&	$0.25_{-0.10}^{+0.12}$	&	$0.17_{-0.07}^{+0.08}$	&	$0.20_{-0.08}^{+0.09}$	&	$0.15_{-0.03}^{+0.03}$	&	1.23	\\
A23	&	26.70	&	$0.01_{-0.01}^{+0.01}$	&	$0.73_{-0.08}^{+0.08}$	&	$4.51_{-1.89}^{+3.68}$	&	$20.92_{-3.30}^{+3.66}$	&	$0.21_{-0.06}^{+0.10}$	&	$0.22_{-0.08}^{+0.10}$	&	$0.21_{-0.07}^{+0.08}$	&	$0.18_{-0.07}^{+0.08}$	&	$0.14_{-0.03}^{+0.04}$	&	0.93	\\
A24	&	27.35	&	$0.01_{-0.01}^{+0.02}$	&	$0.72_{-0.03}^{+0.10}$	&	$5.28_{-2.75}^{+1.81}$	&	$19.53_{-4.25}^{+1.52}$	&	$0.30_{-0.09}^{+0.11}$	&	$0.27_{-0.12}^{+0.09}$	&	$0.25_{-0.06}^{+0.11}$	&	$0.29_{-0.08}^{+0.12}$	&	$0.15_{-0.02}^{+0.06}$	&	1.32	\\
A25	&	27.97	&	$0.01_{-0.01}^{+0.01}$	&	$0.81_{-0.09}^{+0.17}$	&	$2.37_{-1.12}^{+1.71}$	&	$17.78_{-4.15}^{+3.62}$	&	$0.21_{-0.05}^{+0.08}$	&	$0.24_{-0.09}^{+0.09}$	&	$0.25_{-0.07}^{+0.08}$	&	$0.27_{-0.08}^{+0.10}$	&	$0.18_{-0.04}^{+0.07}$	&	1.27	\\
A26	&	28.53	&	$0.01_{-0.01}^{+0.02}$	&	$0.71_{-0.05}^{+0.07}$	&	$5.21_{-2.16}^{+3.56}$	&	$19.38_{-3.12}^{+2.92}$	&	$0.26_{-0.07}^{+0.14}$	&	$0.23_{-0.09}^{+0.11}$	&	$0.27_{-0.07}^{+0.10}$	&	$0.17_{-0.08}^{+0.09}$	&	$0.15_{-0.03}^{+0.04}$	&	1.50	\\
A27	&	29.07	&	$0.01_{-0.01}^{+0.01}$	&	$0.76_{-0.06}^{+0.10}$	&	$3.29_{-1.42}^{+2.34}$	&	$20.38_{-3.70}^{+3.30}$	&	$0.21_{-0.06}^{+0.10}$	&	$0.20_{-0.08}^{+0.08}$	&	$0.15_{-0.06}^{+0.07}$	&	$0.24_{-0.08}^{+0.09}$	&	$0.16_{-0.03}^{+0.04}$	&	1.30	\\
A28	&	29.63	&	$0.01_{-0.01}^{+0.01}$	&	$0.69_{-0.08}^{+0.08}$	&	$2.56_{-0.93}^{+1.82}$	&	$20.79_{-3.57}^{+4.72}$	&	$0.15_{-0.03}^{+0.05}$	&	$0.28_{-0.06}^{+0.07}$	&	$0.22_{-0.06}^{+0.07}$	&	$0.32_{-0.09}^{+0.12}$	&	$0.14_{-0.03}^{+0.03}$	&	1.20	\\
A29	&	30.21	&	$0.01_{-0.01}^{+0.01}$	&	$0.67_{-0.06}^{+0.06}$	&	$3.92_{-1.62}^{+2.67}$	&	$22.47_{-4.12}^{+4.35}$	&	$0.21_{-0.05}^{+0.08}$	&	$0.23_{-0.07}^{+0.07}$	&	$0.15_{-0.06}^{+0.06}$	&	$0.29_{-0.09}^{+0.10}$	&	$0.12_{-0.02}^{+0.03}$	&	1.36	\\
A30	&	30.84	&	$0.01_{-0.01}^{+0.01}$	&	$0.71_{-0.07}^{+0.12}$	&	$2.76_{-1.13}^{+1.42}$	&	$18.34_{-3.99}^{+3.07}$	&	$0.21_{-0.05}^{+0.09}$	&	$0.29_{-0.07}^{+0.08}$	&	$0.23_{-0.06}^{+0.09}$	&	$0.28_{-0.09}^{+0.13}$	&	$0.15_{-0.02}^{+0.06}$	&	1.59	\\
A31	&	31.54	&	$0.01_{-0.01}^{+0.01}$	&	$0.83_{-0.15}^{+0.24}$	&	$1.28_{-0.62}^{+0.69}$	&	$16.53_{-4.77}^{+3.40}$	&	$0.14_{-0.02}^{+0.03}$	&	$0.23_{-0.07}^{+0.05}$	&	$0.24_{-0.07}^{+0.08}$	&	$0.22_{-0.09}^{+0.09}$	&	$0.19_{-0.04}^{+0.08}$	&	1.26	\\
\hline
B01 (*)	&	3.46	&	$0.09_{-0.06}^{+0.05}$	&	$1.11_{-0.18}^{+0.17}$	&	$6.52_{-2.85}^{+6.62}$	&	$17.58_{-1.58}^{+1.40}$	&	$0.48_{-0.48}^{+0.35}$	&	$0.67_{-0.62}^{+1.94}$	&	$1.04_{-0.46}^{+1.28}$	&	$0.96_{-0.39}^{+0.96}$	&	$0.31_{-0.16}^{+0.26}$	&	0.93	\\
B02 (*)	&	6.35	&	$0.01_{-0.01}^{+0.03}$	&	$0.74_{-0.06}^{+0.04}$	&	$7.47_{-1.66}^{+5.46}$	&	$19.58_{-2.46}^{+3.40}$	&	$0.28_{-0.08}^{+0.07}$	&	$0.40_{-0.11}^{+0.10}$	&	$0.32_{-0.10}^{+0.09}$	&	$0.45_{-0.12}^{+0.12}$	&	$0.15_{-0.04}^{+0.03}$	&	1.46	\\
B03 (*)	&	8.20	&	$0.01_{-0.01}^{+0.01}$	&	$0.75_{-0.06}^{+0.07}$	&	$7.96_{-1.57}^{+2.42}$	&	$20.67_{-2.55}^{+2.53}$	&	$0.25_{-0.09}^{+0.11}$	&	$0.28_{-0.13}^{+0.16}$	&	$0.27_{-0.10}^{+0.16}$	&	$0.39_{-0.12}^{+0.18}$	&	$0.16_{-0.04}^{+0.06}$	&	1.15	\\
B04 (*)	&	9.74	&	$0.04_{-0.04}^{+0.05}$	&	$0.67_{-0.06}^{+0.08}$	&	$6.29_{-0.93}^{+1.26}$	&	$26.44_{-1.90}^{+6.78}$	&	$0.18_{-0.05}^{+0.10}$	&	$0.23_{-0.09}^{+0.08}$	&	$0.12_{-0.08}^{+0.08}$	&	$0.32_{-0.05}^{+0.11}$	&	$0.10_{-0.03}^{+0.04}$	&	1.20	\\
B05 (*)	&	11.15	&	$0.01_{-0.01}^{+0.01}$	&	$0.63_{-0.03}^{+0.04}$	&	$13.14_{-3.07}^{+15.16}$	&	$26.94_{-4.42}^{+4.42}$	&	$0.33_{-0.08}^{+0.10}$	&	$0.25_{-0.08}^{+0.09}$	&	$0.16_{-0.07}^{+0.08}$	&	$0.32_{-0.09}^{+0.11}$	&	$0.11_{-0.02}^{+0.02}$	&	1.59	\\
B06 (*)	&	12.55	&	$0.01_{-0.01}^{+0.03}$	&	$0.72_{-0.04}^{+0.04}$	&	$12.45_{-2.11}^{+3.01}$	&	$22.95_{-1.77}^{+2.66}$	&	$0.35_{-0.10}^{+0.11}$	&	$0.29_{-0.11}^{+0.10}$	&	$0.23_{-0.09}^{+0.11}$	&	$0.36_{-0.10}^{+0.11}$	&	$0.12_{-0.02}^{+0.03}$	&	1.13	\\
B07	&	13.94	&	$0.01_{-0.01}^{+0.01}$	&	$0.71_{-0.05}^{+0.04}$	&	$13.27_{-5.69}^{+9.40}$	&	$21.61_{-2.76}^{+2.77}$	&	$0.45_{-0.14}^{+0.20}$	&	$0.47_{-0.15}^{+0.17}$	&	$0.15_{-0.07}^{+0.08}$	&	$0.30_{-0.09}^{+0.11}$	&	$0.11_{-0.02}^{+0.02}$	&	1.59	\\
B08	&	15.21	&	$0.03_{-0.03}^{+0.06}$	&	$0.83_{-0.07}^{+0.42}$	&	$8.36_{-7.21}^{+12.13}$	&	$20.90_{-3.22}^{+3.96}$	&	$0.41_{-0.27}^{+0.24}$	&	$0.24_{-0.24}^{+0.26}$	&	$0.08_{-0.08}^{+0.07}$	&	$0.22_{-0.08}^{+0.10}$	&	$0.13_{-0.03}^{+0.14}$	&	1.51	\\
B09	&	16.41	&	$0.13_{-0.05}^{+0.06}$	&	$1.14_{-0.22}^{+0.21}$	&	$0.77_{-0.27}^{+0.70}$	&	$18.51_{-3.49}^{+6.29}$	&	$0.10_{-0.03}^{+0.03}$	&	$0.14_{-0.05}^{+0.05}$	&	$0.08_{-0.05}^{+0.05}$	&	$0.17_{-0.06}^{+0.07}$	&	$0.14_{-0.04}^{+0.04}$	&	1.59	\\
B10	&	17.54	&	$0.01_{-0.01}^{+0.05}$	&	$0.73_{-0.04}^{+0.04}$	&	$17.77_{-10.15}^{+14.80}$	&	$22.65_{-2.93}^{+5.13}$	&	$0.43_{-0.21}^{+0.21}$	&	$0.44_{-0.20}^{+0.19}$	&	$0.21_{-0.10}^{+0.11}$	&	$0.36_{-0.11}^{+0.12}$	&	$0.13_{-0.03}^{+0.03}$	&	1.59	\\
B11	&	18.48	&	$0.01_{-0.01}^{+0.05}$	&	$0.75_{-0.04}^{+0.04}$	&	$15.25_{-9.36}^{+7.18}$	&	$16.19_{-1.85}^{+3.81}$	&	$0.65_{-0.34}^{+0.38}$	&	$0.51_{-0.28}^{+0.22}$	&	$0.29_{-0.13}^{+0.06}$	&	$0.34_{-0.12}^{+0.11}$	&	$0.21_{-0.05}^{+0.04}$	&	1.47	\\
B12	&	19.35	&	$0.01_{-0.01}^{+0.03}$	&	$0.70_{-0.03}^{+0.04}$	&	$16.78_{-8.22}^{+12.22}$	&	$21.85_{-3.53}^{+2.32}$	&	$0.46_{-0.15}^{+0.25}$	&	$0.41_{-0.15}^{+0.19}$	&	$0.18_{-0.07}^{+0.11}$	&	$0.27_{-0.08}^{+0.11}$	&	$0.15_{-0.02}^{+0.04}$	&	1.18	\\
B13	&	20.30	&	$0.03_{-0.03}^{+0.06}$	&	$0.70_{-0.05}^{+0.04}$	&	$9.21_{-5.61}^{+10.45}$	&	$22.12_{-4.34}^{+4.97}$	&	$0.45_{-0.24}^{+0.35}$	&	$0.26_{-0.12}^{+0.20}$	&	$0.10_{-0.07}^{+0.10}$	&	$0.23_{-0.08}^{+0.10}$	&	$0.15_{-0.03}^{+0.04}$	&	1.17	\\
B14	&	21.38	&	$0.01_{-0.01}^{+0.01}$	&	$0.64_{-0.04}^{+0.04}$	&	$12.62_{-5.02}^{+8.97}$	&	$22.26_{-2.97}^{+3.28}$	&	$0.44_{-0.14}^{+0.20}$	&	$0.42_{-0.12}^{+0.15}$	&	$0.16_{-0.07}^{+0.08}$	&	$0.32_{-0.10}^{+0.11}$	&	$0.15_{-0.02}^{+0.03}$	&	1.17	\\
B15	&	22.56	&	$0.01_{-0.01}^{+0.07}$	&	$0.63_{-0.06}^{+0.05}$	&	$5.76_{-3.25}^{+2.76}$	&	$22.01_{-3.29}^{+7.73}$	&	$0.27_{-0.14}^{+0.15}$	&	$0.27_{-0.10}^{+0.11}$	&	$0.18_{-0.08}^{+0.09}$	&	$0.23_{-0.09}^{+0.11}$	&	$0.13_{-0.03}^{+0.03}$	&	1.49	\\
B16	&	23.79	&	$0.01_{-0.01}^{+0.01}$	&	$0.71_{-0.08}^{+0.09}$	&	$3.19_{-1.3}^{+2.29}$	&	$19.48_{-3.42}^{+4.21}$	&	$0.21_{-0.05}^{+0.08}$	&	$0.24_{-0.07}^{+0.08}$	&	$0.15_{-0.06}^{+0.07}$	&	$0.27_{-0.09}^{+0.11}$	&	$0.13_{-0.03}^{+0.04}$	&	1.45	\\
B17	&	24.96	&	$0.01_{-0.01}^{+0.01}$	&	$0.76_{-0.07}^{+0.09}$	&	$4.08_{-1.05}^{+3.59}$	&	$21.26_{-4.33}^{+3.69}$	&	$0.28_{-0.08}^{+0.14}$	&	$0.16_{-0.08}^{+0.09}$	&	$0.08_{-0.06}^{+0.06}$	&	$0.15_{-0.07}^{+0.08}$	&	$0.13_{-0.03}^{+0.05}$	&	1.34	\\
B18	&	25.91	&	$0.01_{-0.01}^{+0.01}$	&	$0.87_{-0.14}^{+0.27}$	&	$1.65_{-0.86}^{+1.37}$	&	$14.37_{-4.29}^{+2.99}$	&	$0.16_{-0.04}^{+0.06}$	&	$0.16_{-0.04}^{+0.08}$	&	$0.14_{-0.06}^{+0.07}$	&	$0.22_{-0.09}^{+0.10}$	&	$0.18_{-0.05}^{+0.09}$	&	1.48	\\
B19	&	26.68	&	$0.01_{-0.01}^{+0.01}$	&	$0.79_{-0.03}^{+0.18}$	&	$1.82_{-0.79}^{+1.03}$	&	$19.37_{-4.72}^{+1.65}$	&	$0.15_{-0.03}^{+0.05}$	&	$0.20_{-0.07}^{+0.05}$	&	$0.19_{-0.06}^{+0.07}$	&	$0.19_{-0.08}^{+0.08}$	&	$0.16_{-0.03}^{+0.07}$	&	1.43	\\
B20	&	27.31	&	$0.01_{-0.01}^{+0.01}$	&	$0.84_{-0.10}^{+0.20}$	&	$1.42_{-0.66}^{+0.92}$	&	$14.94_{-3.89}^{+3.37}$	&	$0.16_{-0.03}^{+0.05}$	&	$0.18_{-0.08}^{+0.07}$	&	$0.12_{-0.06}^{+0.07}$	&	$0.28_{-0.09}^{+0.11}$	&	$0.20_{-0.05}^{+0.08}$	&	1.48	\\
B21	&	27.83	&	$0.01_{-0.01}^{+0.01}$	&	$0.69_{-0.08}^{+0.11}$	&	$2.17_{-0.93}^{+1.68}$	&	$19.39_{-4.05}^{+4.82}$	&	$0.15_{-0.03}^{+0.05}$	&	$0.18_{-0.05}^{+0.06}$	&	$0.16_{-0.06}^{+0.06}$	&	$0.15_{-0.08}^{+0.09}$	&	$0.13_{-0.03}^{+0.04}$	&	1.33	\\
B22	&	28.38	&	$0.01_{-0.01}^{+0.01}$	&	$0.68_{-0.08}^{+0.15}$	&	$1.38_{-0.65}^{+0.88}$	&	$15.28_{-4.16}^{+3.99}$	&	$0.17_{-0.03}^{+0.04}$	&	$0.25_{-0.06}^{+0.06}$	&	$0.24_{-0.07}^{+0.10}$	&	$0.28_{-0.11}^{+0.12}$	&	$0.18_{-0.04}^{+0.07}$	&	0.97	\\
    \hline
    \end{tabular}%
  \label{tab:six_directions}%
\end{table*}
    
\begin{table*}
\contcaption{}
\setlength{\tabcolsep}{3.5pt}
\renewcommand{\arraystretch}{1.5}
\scriptsize
 \centering
    \begin{tabular}{cccccccccccc}
\hline
Region &   $r$    & $n_{\rm H}$            & $kT$                   & $n_{\rm e}t$             & $EM$                   & O                      & Ne                     & Mg                     & Si                     & Fe         & $\chi^{2}_{\nu}$\\
       &  ($''$)  & ($10^{22}$cm$^{-2}$)   & (keV)                  & ($10^{11}$cm$^{-3}$s)    & ($10^{57}$cm$^{-3}$)   &                        &                        &                        &                        &                        & \\
\hline
C01 (*)	&	3.51	&	$0.01_{-0.01}^{+0.01}$	&	$0.80_{-0.05}^{+0.06}$	&	$6.79_{-0.86}^{+2.34}$	&	$16.32_{-1.93}^{+2.90}$	&	$0.54_{-0.11}^{+0.15}$	&	$0.37_{-0.16}^{+0.11}$	&	$0.37_{-0.11}^{+0.16}$	&	$0.37_{-0.11}^{+0.15}$	&	$0.19_{-0.04}^{+0.06}$	&	1.59	\\
C02 (*)	&	6.49	&	$0.01_{-0.01}^{+0.02}$	&	$0.67_{-0.04}^{+0.05}$	&	$9.65_{-3.91}^{+28.45}$	&	$23.69_{-5.84}^{+7.01}$	&	$0.22_{-0.10}^{+0.11}$	&	$0.34_{-0.10}^{+0.13}$	&	$0.22_{-0.09}^{+0.08}$	&	$0.35_{-0.11}^{+0.15}$	&	$0.11_{-0.02}^{+0.04}$	&	1.01	\\
C03 (*)	&	8.34	&	$0.01_{-0.01}^{+0.01}$	&	$0.77_{-0.08}^{+0.15}$	&	$8.47_{-4.07}^{+33.52}$	&	$19.54_{-3.11}^{+2.81}$	&	$0.23_{-0.07}^{+0.14}$	&	$0.23_{-0.23}^{+0.12}$	&	$0.30_{-0.10}^{+0.17}$	&	$0.27_{-0.11}^{+0.14}$	&	$0.21_{-0.06}^{+0.18}$	&	1.51	\\
C04 (*)	&	10.01	&	$0.03_{-0.03}^{+0.05}$	&	$0.79_{-0.07}^{+0.08}$	&	$8.57_{-4.63}^{+8.57}$	&	$24.62_{-6.67}^{+7.59}$	&	$0.22_{-0.22}^{+0.17}$	&	$0.30_{-0.19}^{+0.17}$	&	$0.19_{-0.11}^{+0.14}$	&	$0.14_{-0.10}^{+0.11}$	&	$0.14_{-0.05}^{+0.05}$	&	1.36	\\
C05 (*)	&	11.66	&	$0.01_{-0.01}^{+0.05}$	&	$0.84_{-0.07}^{+0.07}$	&	$17.95_{-8.47}^{+449.42}$	&	$20.11_{-2.40}^{+2.32}$	&	$0.44_{-0.35}^{+0.21}$	&	$0.14_{-0.14}^{+0.29}$	&	$0.18_{-0.15}^{+0.17}$	&	$0.24_{-0.13}^{+0.15}$	&	$0.17_{-0.05}^{+0.07}$	&	1.00	\\
C06 (*)	&	12.97	&	$0.01_{-0.01}^{+0.03}$	&	$0.66_{-0.06}^{+0.04}$	&	$8.37_{-2.18}^{+12.12}$	&	$22.87_{-5.35}^{+5.77}$	&	$0.33_{-0.11}^{+0.07}$	&	$0.26_{-0.10}^{+0.07}$	&	$0.17_{-0.09}^{+0.07}$	&	$0.23_{-0.11}^{+0.10}$	&	$0.13_{-0.03}^{+0.02}$	&	1.06	\\
C07	&	14.06	&	$0.01_{-0.01}^{+0.01}$	&	$0.62_{-0.05}^{+0.07}$	&	$5.15_{-2.15}^{+4.04}$	&	$27.42_{-4.64}^{+4.64}$	&	$0.16_{-0.04}^{+0.06}$	&	$0.22_{-0.06}^{+0.06}$	&	$0.14_{-0.06}^{+0.06}$	&	$0.18_{-0.07}^{+0.08}$	&	$0.10_{-0.02}^{+0.02}$	&	1.22	\\
C08	&	15.04	&	$0.01_{-0.01}^{+0.01}$	&	$0.72_{-0.03}^{+0.07}$	&	$5.53_{-2.46}^{+4.47}$	&	$22.70_{-3.44}^{+3.57}$	&	$0.22_{-0.07}^{+0.12}$	&	$0.24_{-0.09}^{+0.10}$	&	$0.10_{-0.06}^{+0.06}$	&	$0.18_{-0.07}^{+0.08}$	&	$0.13_{-0.03}^{+0.04}$	&	1.51	\\
C09	&	15.92	&	$0.01_{-0.01}^{+0.01}$	&	$0.71_{-0.06}^{+0.06}$	&	$6.16_{-2.48}^{+4.01}$	&	$22.10_{-2.99}^{+3.36}$	&	$0.30_{-0.09}^{+0.13}$	&	$0.34_{-0.11}^{+0.13}$	&	$0.23_{-0.07}^{+0.08}$	&	$0.28_{-0.09}^{+0.10}$	&	$0.14_{-0.03}^{+0.03}$	&	1.23	\\
C10	&	16.71	&	$0.01_{-0.01}^{+0.01}$	&	$0.72_{-0.06}^{+0.08}$	&	$6.41_{-3.02}^{+6.34}$	&	$23.36_{-3.61}^{+4.04}$	&	$0.19_{-0.06}^{+0.11}$	&	$0.25_{-0.10}^{+0.12}$	&	$0.17_{-0.06}^{+0.07}$	&	$0.20_{-0.08}^{+0.08}$	&	$0.11_{-0.02}^{+0.03}$	&	1.27	\\
C11	&	17.40	&	$0.01_{-0.01}^{+0.01}$	&	$0.73_{-0.04}^{+0.05}$	&	$11.02_{-5.18}^{+10.96}$	&	$22.28_{-2.62}^{+2.90}$	&	$0.36_{-0.14}^{+0.21}$	&	$0.28_{-0.13}^{+0.16}$	&	$0.15_{-0.07}^{+0.08}$	&	$0.21_{-0.08}^{+0.09}$	&	$0.14_{-0.02}^{+0.03}$	&	1.15	\\
C12	&	18.03	&	$0.01_{-0.01}^{+0.01}$	&	$0.78_{-0.06}^{+0.19}$	&	$3.86_{-2.49}^{+4.37}$	&	$19.01_{-5.20}^{+2.96}$	&	$0.19_{-0.07}^{+0.14}$	&	$0.21_{-0.11}^{+0.12}$	&	$0.18_{-0.07}^{+0.07}$	&	$0.22_{-0.08}^{+0.09}$	&	$0.16_{-0.03}^{+0.07}$	&	1.46	\\
C13	&	18.60	&	$0.01_{-0.01}^{+0.01}$	&	$0.70_{-0.07}^{+0.06}$	&	$4.39_{-1.99}^{+4.89}$	&	$20.06_{-3.05}^{+4.08}$	&	$0.16_{-0.05}^{+0.10}$	&	$0.26_{-0.08}^{+0.12}$	&	$0.15_{-0.06}^{+0.07}$	&	$0.26_{-0.09}^{+0.11}$	&	$0.16_{-0.03}^{+0.04}$	&	1.42	\\
C14	&	19.11	&	$0.01_{-0.01}^{+0.01}$	&	$0.71_{-0.05}^{+0.07}$	&	$4.93_{-2.18}^{+4.38}$	&	$21.21_{-3.18}^{+3.21}$	&	$0.20_{-0.06}^{+0.12}$	&	$0.23_{-0.09}^{+0.10}$	&	$0.15_{-0.06}^{+0.07}$	&	$0.20_{-0.08}^{+0.08}$	&	$0.16_{-0.01}^{+0.04}$	&	1.23	\\
C15	&	19.59	&	$0.01_{-0.01}^{+0.01}$	&	$0.78_{-0.03}^{+0.11}$	&	$3.12_{-1.54}^{+1.67}$	&	$20.79_{-4.40}^{+3.79}$	&	$0.16_{-0.05}^{+0.09}$	&	$0.13_{-0.07}^{+0.07}$	&	$0.14_{-0.06}^{+0.06}$	&	$0.15_{-0.07}^{+0.08}$	&	$0.16_{-0.03}^{+0.05}$	&	1.39	\\
C16	&	20.07	&	$0.01_{-0.01}^{+0.01}$	&	$0.71_{-0.04}^{+0.05}$	&	$7.40_{-3.35}^{+6.26}$	&	$24.41_{-3.21}^{+3.36}$	&	$0.25_{-0.09}^{+0.13}$	&	$0.18_{-0.08}^{+0.10}$	&	$0.13_{-0.06}^{+0.07}$	&	$0.12_{-0.07}^{+0.07}$	&	$0.15_{-0.02}^{+0.03}$	&	1.33	\\
C17	&	20.54	&	$0.01_{-0.01}^{+0.01}$	&	$1.00_{-0.19}^{+0.39}$	&	$1.05_{-0.54}^{+0.94}$	&	$13.78_{-4.26}^{+4.13}$	&	$0.10_{-0.02}^{+0.03}$	&	$0.05_{-0.05}^{+0.08}$	&	$0.19_{-0.07}^{+0.08}$	&	$0.12_{-0.07}^{+0.08}$	&	$0.27_{-0.07}^{+0.11}$	&	0.86	\\
C18	&	20.99	&	$0.01_{-0.01}^{+0.01}$	&	$0.72_{-0.08}^{+0.10}$	&	$2.27_{-0.91}^{+1.83}$	&	$20.65_{-3.81}^{+2.35}$	&	$0.10_{-0.02}^{+0.04}$	&	$0.17_{-0.06}^{+0.06}$	&	$0.17_{-0.06}^{+0.06}$	&	$0.20_{-0.08}^{+0.08}$	&	$0.16_{-0.03}^{+0.03}$	&	1.59	\\
C19	&	21.42	&	$0.01_{-0.01}^{+0.01}$	&	$0.65_{-0.02}^{+0.05}$	&	$9.48_{-4.02}^{+7.62}$	&	$20.77_{-2.93}^{+3.14}$	&	$0.29_{-0.10}^{+0.16}$	&	$0.29_{-0.10}^{+0.13}$	&	$0.23_{-0.08}^{+0.09}$	&	$0.29_{-0.09}^{+0.10}$	&	$0.17_{-0.03}^{+0.03}$	&	1.34	\\
C20	&	21.84	&	$0.01_{-0.01}^{+0.01}$	&	$0.64_{-0.05}^{+0.08}$	&	$3.55_{-1.45}^{+2.20}$	&	$22.47_{-3.98}^{+3.72}$	&	$0.13_{-0.03}^{+0.05}$	&	$0.19_{-0.06}^{+0.06}$	&	$0.18_{-0.06}^{+0.07}$	&	$0.25_{-0.09}^{+0.10}$	&	$0.16_{-0.03}^{+0.04}$	&	1.20	\\
C21	&	22.26	&	$0.01_{-0.01}^{+0.01}$	&	$0.66_{-0.05}^{+0.10}$	&	$2.82_{-1.25}^{+1.75}$	&	$20.07_{-4.23}^{+3.68}$	&	$0.15_{-0.03}^{+0.06}$	&	$0.20_{-0.10}^{+0.07}$	&	$0.14_{-0.06}^{+0.07}$	&	$0.18_{-0.08}^{+0.09}$	&	$0.18_{-0.03}^{+0.05}$	&	1.22	\\
C22	&	22.68	&	$0.01_{-0.01}^{+0.01}$	&	$0.65_{-0.05}^{+0.06}$	&	$4.32_{-1.77}^{+2.81}$	&	$21.43_{-3.38}^{+3.50}$	&	$0.20_{-0.06}^{+0.09}$	&	$0.20_{-0.07}^{+0.08}$	&	$0.19_{-0.07}^{+0.08}$	&	$0.14_{-0.08}^{+0.09}$	&	$0.16_{-0.03}^{+0.04}$	&	1.38	\\
C23	&	23.10	&	$0.01_{-0.01}^{+0.01}$	&	$0.65_{-0.05}^{+0.10}$	&	$2.63_{-1.22}^{+1.60}$	&	$19.47_{-4.28}^{+3.58}$	&	$0.15_{-0.03}^{+0.05}$	&	$0.18_{-0.07}^{+0.06}$	&	$0.21_{-0.07}^{+0.08}$	&	$0.19_{-0.09}^{+0.10}$	&	$0.18_{-0.03}^{+0.05}$	&	1.23	\\
C24	&	23.55	&	$0.01_{-0.01}^{+0.01}$	&	$0.61_{-0.05}^{+0.07}$	&	$2.47_{-0.91}^{+1.34}$	&	$21.59_{-3.85}^{+3.94}$	&	$0.12_{-0.03}^{+0.04}$	&	$0.24_{-0.06}^{+0.06}$	&	$0.25_{-0.07}^{+0.08}$	&	$0.27_{-0.10}^{+0.12}$	&	$0.17_{-0.03}^{+0.04}$	&	1.25	\\
C25	&	24.04	&	$0.01_{-0.01}^{+0.01}$	&	$0.62_{-0.06}^{+0.09}$	&	$2.33_{-0.92}^{+1.34}$	&	$19.84_{-3.94}^{+4.10}$	&	$0.14_{-0.03}^{+0.04}$	&	$0.24_{-0.06}^{+0.06}$	&	$0.19_{-0.07}^{+0.07}$	&	$0.25_{-0.10}^{+0.12}$	&	$0.17_{-0.03}^{+0.04}$	&	1.48	\\
C26	&	24.61	&	$0.01_{-0.01}^{+0.01}$	&	$0.58_{-0.03}^{+0.09}$	&	$2.37_{-1.00}^{+1.17}$	&	$21.86_{-5.02}^{+3.10}$	&	$0.17_{-0.03}^{+0.04}$	&	$0.27_{-0.06}^{+0.06}$	&	$0.27_{-0.07}^{+0.08}$	&	$0.27_{-0.11}^{+0.11}$	&	$0.15_{-0.02}^{+0.04}$	&	1.42	\\
\hline  
D01 (*)	&	3.36	&	$0.01_{-0.01}^{+0.01}$	&	$0.77_{-0.08}^{+0.05}$	&	$12.45_{-5.39}^{+12.45}$	&	$18.89_{-3.76}^{+5.14}$	&	$0.29_{-0.10}^{+0.12}$	&	$0.35_{-0.09}^{+0.16}$	&	$0.28_{-0.14}^{+0.10}$	&	$0.30_{-0.13}^{+0.09}$	&	$0.17_{-0.05}^{+0.03}$	&	1.14	\\
D02 (*)	&	6.36	&	$0.01_{-0.01}^{+0.01}$	&	$0.77_{-0.05}^{+0.06}$	&	$7.16_{-1.45}^{+2.25}$	&	$20.19_{-2.42}^{+2.53}$	&	$0.20_{-0.09}^{+0.09}$	&	$0.33_{-0.14}^{+0.15}$	&	$0.25_{-0.09}^{+0.15}$	&	$0.26_{-0.10}^{+0.14}$	&	$0.17_{-0.04}^{+0.06}$	&	0.97	\\
D03 (*)	&	8.56	&	$0.01_{-0.01}^{+0.01}$	&	$0.73_{-0.11}^{+0.08}$	&	$7.46_{-2.05}^{+3.97}$	&	$22.13_{-1.97}^{+2.78}$	&	$0.14_{-0.10}^{+0.09}$	&	$0.32_{-0.11}^{+0.14}$	&	$0.16_{-0.09}^{+0.13}$	&	$0.20_{-0.10}^{+0.12}$	&	$0.13_{-0.03}^{+0.05}$	&	1.19	\\
D04 (*)	&	10.46	&	$0.01_{-0.01}^{+0.01}$	&	$0.77_{-0.08}^{+0.11}$	&	$8.58_{-2.82}^{+6.44}$	&	$19.36_{-3.11}^{+2.73}$	&	$0.14_{-0.14}^{+0.12}$	&	$0.59_{-0.19}^{+0.30}$	&	$0.36_{-0.14}^{+0.31}$	&	$0.36_{-0.15}^{+0.22}$	&	$0.18_{-0.06}^{+0.07}$	&	1.21	\\
D05 (*)	&	12.26	&	$0.01_{-0.01}^{+0.01}$	&	$0.74_{-0.06}^{+0.07}$	&	$6.16_{-0.82}^{+1.06}$	&	$20.79_{-1.45}^{+2.74}$	&	$0.26_{-0.06}^{+0.06}$	&	$0.29_{-0.09}^{+0.08}$	&	$0.19_{-0.07}^{+0.05}$	&	$0.27_{-0.09}^{+0.07}$	&	$0.13_{-0.03}^{+0.04}$	&	1.21	\\
D06 (*)	&	13.86	&	$0.01_{-0.01}^{+0.03}$	&	$0.73_{-0.05}^{+0.06}$	&	$11.14_{-3.07}^{+5.99}$	&	$20.33_{-2.29}^{+2.96}$	&	$0.23_{-0.14}^{+0.14}$	&	$0.49_{-0.17}^{+0.24}$	&	$0.26_{-0.12}^{+0.18}$	&	$0.24_{-0.11}^{+0.15}$	&	$0.17_{-0.04}^{+0.06}$	&	1.31	\\
D07	&	15.24	&	$0.01_{-0.01}^{+0.01}$	&	$0.80_{-0.08}^{+0.22}$	&	$2.27_{-1.27}^{+1.89}$	&	$18.56_{-5.34}^{+3.84}$	&	$0.15_{-0.04}^{+0.07}$	&	$0.19_{-0.08}^{+0.07}$	&	$0.16_{-0.06}^{+0.07}$	&	$0.16_{-0.07}^{+0.08}$	&	$0.15_{-0.03}^{+0.04}$	&	1.20	\\
D08	&	16.38	&	$0.01_{-0.01}^{+0.01}$	&	$0.79_{-0.08}^{+0.23}$	&	$2.74_{-0.82}^{+2.52}$	&	$18.65_{-5.52}^{+3.47}$	&	$0.17_{-0.05}^{+0.10}$	&	$0.21_{-0.11}^{+0.09}$	&	$0.16_{-0.06}^{+0.07}$	&	$0.18_{-0.08}^{+0.09}$	&	$0.17_{-0.03}^{+0.09}$	&	1.59	\\
D09	&	17.34	&	$0.01_{-0.01}^{+0.01}$	&	$0.77_{-0.07}^{+0.14}$	&	$2.43_{-1.07}^{+1.39}$	&	$17.62_{-3.78}^{+3.09}$	&	$0.16_{-0.04}^{+0.06}$	&	$0.28_{-0.09}^{+0.08}$	&	$0.19_{-0.07}^{+0.07}$	&	$0.28_{-0.09}^{+0.10}$	&	$0.18_{-0.04}^{+0.06}$	&	1.59	\\
D10	&	18.18	&	$0.01_{-0.01}^{+0.01}$	&	$0.77_{-0.07}^{+0.12}$	&	$2.46_{-1.10}^{+2.05}$	&	$17.65_{-3.66}^{+3.34}$	&	$0.14_{-0.04}^{+0.06}$	&	$0.18_{-0.07}^{+0.07}$	&	$0.17_{-0.06}^{+0.07}$	&	$0.24_{-0.09}^{+0.10}$	&	$0.17_{-0.03}^{+0.05}$	&	1.11	\\
D11	&	18.96	&	$0.01_{-0.01}^{+0.01}$	&	$0.71_{-0.05}^{+0.16}$	&	$2.52_{-1.25}^{+1.16}$	&	$20.38_{-5.34}^{+2.04}$	&	$0.11_{-0.02}^{+0.04}$	&	$0.22_{-0.09}^{+0.05}$	&	$0.19_{-0.07}^{+0.07}$	&	$0.28_{-0.11}^{+0.09}$	&	$0.15_{-0.02}^{+0.07}$	&	1.46	\\
D12	&	19.75	&	$0.01_{-0.01}^{+0.01}$	&	$0.79_{-0.06}^{+0.23}$	&	$2.19_{-1.11}^{+1.61}$	&	$16.34_{-4.62}^{+3.83}$	&	$0.16_{-0.04}^{+0.09}$	&	$0.22_{-0.09}^{+0.07}$	&	$0.19_{-0.08}^{+0.07}$	&	$0.23_{-0.05}^{+0.09}$	&	$0.20_{-0.04}^{+0.11}$	&	1.59	\\
D13	&	20.55	&	$0.01_{-0.01}^{+0.01}$	&	$0.71_{-0.07}^{+0.10}$	&	$2.79_{-1.13}^{+1.88}$	&	$20.03_{-3.75}^{+4.97}$	&	$0.15_{-0.04}^{+0.05}$	&	$0.19_{-0.04}^{+0.07}$	&	$0.18_{-0.06}^{+0.07}$	&	$0.22_{-0.08}^{+0.09}$	&	$0.16_{-0.03}^{+0.04}$	&	1.56	\\
D14	&	21.32	&	$0.01_{-0.01}^{+0.01}$	&	$0.69_{-0.10}^{+0.07}$	&	$2.68_{-0.90}^{+2.45}$	&	$18.85_{-2.88}^{+5.37}$	&	$0.15_{-0.04}^{+0.06}$	&	$0.26_{-0.06}^{+0.09}$	&	$0.28_{-0.07}^{+0.09}$	&	$0.28_{-0.09}^{+0.13}$	&	$0.15_{-0.04}^{+0.03}$	&	1.12	\\
D15	&	22.06	&	$0.01_{-0.01}^{+0.01}$	&	$0.66_{-0.06}^{+0.11}$	&	$2.00_{-0.86}^{+1.24}$	&	$21.02_{-4.66}^{+4.21}$	&	$0.10_{-0.02}^{+0.03}$	&	$0.19_{-0.05}^{+0.05}$	&	$0.15_{-0.06}^{+0.06}$	&	$0.22_{-0.09}^{+0.09}$	&	$0.15_{-0.03}^{+0.04}$	&	1.24	\\
D16	&	22.78	&	$0.01_{-0.01}^{+0.01}$	&	$0.63_{-0.06}^{+0.11}$	&	$2.05_{-0.89}^{+1.15}$	&	$23.83_{-5.55}^{+4.81}$	&	$0.10_{-0.02}^{+0.02}$	&	$0.22_{-0.04}^{+0.04}$	&	$0.12_{-0.05}^{+0.05}$	&	$0.16_{-0.08}^{+0.09}$	&	$0.11_{-0.02}^{+0.03}$	&	1.26	\\
D17	&	23.47	&	$0.01_{-0.01}^{+0.01}$	&	$0.61_{-0.05}^{+0.08}$	&	$2.87_{-1.11}^{+1.76}$	&	$23.10_{-4.30}^{+4.11}$	&	$0.15_{-0.03}^{+0.04}$	&	$0.22_{-0.05}^{+0.05}$	&	$0.15_{-0.06}^{+0.06}$	&	$0.21_{-0.09}^{+0.11}$	&	$0.13_{-0.02}^{+0.03}$	&	1.16	\\
D18	&	24.11	&	$0.01_{-0.01}^{+0.01}$	&	$0.67_{-0.08}^{+0.14}$	&	$1.46_{-0.66}^{+0.92}$	&	$20.91_{-5.32}^{+5.05}$	&	$0.12_{-0.02}^{+0.02}$	&	$0.23_{-0.05}^{+0.05}$	&	$0.12_{-0.05}^{+0.06}$	&	$0.21_{-0.09}^{+0.09}$	&	$0.14_{-0.03}^{+0.03}$	&	1.59	\\
D19	&	24.66	&	$0.01_{-0.01}^{+0.02}$	&	$0.67_{-0.09}^{+0.10}$	&	$1.92_{-0.75}^{+1.76}$	&	$16.9_{-3.33}^{+5.13}$	&	$0.15_{-0.03}^{+0.05}$	&	$0.23_{-0.05}^{+0.07}$	&	$0.20_{-0.07}^{+0.07}$	&	$0.27_{-0.10}^{+0.13}$	&	$0.15_{-0.04}^{+0.04}$	&	1.25	\\
D20	&	25.17	&	$0.01_{-0.01}^{+0.01}$	&	$0.62_{-0.03}^{+0.06}$	&	$2.14_{-0.94}^{+1.17}$	&	$19.59_{-4.51}^{+3.96}$	&	$0.16_{-0.03}^{+0.04}$	&	$0.25_{-0.07}^{+0.06}$	&	$0.24_{-0.07}^{+0.08}$	&	$0.35_{-0.12}^{+0.14}$	&	$0.16_{-0.03}^{+0.03}$	&	1.57	\\
D21	&	25.75	&	$0.01_{-0.01}^{+0.01}$	&	$0.56_{-0.07}^{+0.08}$	&	$2.33_{-0.97}^{+1.89}$	&	$22.01_{-4.79}^{+5.83}$	&	$0.17_{-0.03}^{+0.04}$	&	$0.29_{-0.03}^{+0.06}$	&	$0.20_{-0.06}^{+0.07}$	&	$0.39_{-0.13}^{+0.16}$	&	$0.13_{-0.02}^{+0.03}$	&	1.28	\\
\hline  
   \end{tabular}%
\end{table*}

\begin{table*}
\contcaption{}
\setlength{\tabcolsep}{3.5pt}
\scriptsize
\renewcommand{\arraystretch}{1.5}
 \centering
    \begin{tabular}{cccccccccccc}
\hline
Region &   $r$    & $n_{\rm H}$            & $kT$                   & $n_{\rm e}t$             & $EM$                   & O                      & Ne                     & Mg                     & Si                     & Fe         & $\chi^{2}_{\nu}$\\
       &  ($''$)  & ($10^{22}$cm$^{-2}$)   & (keV)                  & ($10^{11}$cm$^{-3}$s)    & ($10^{57}$cm$^{-3}$)   &                        &                        &                        &                        &                        & \\
\hline
E01 (*)	&	3.26	&	$0.01_{-0.01}^{+0.01}$	&	$0.67_{-0.06}^{+0.04}$	&	$7.76_{-1.48}^{+5.31}$	&	$20.42_{-2.90}^{+3.81}$	&	$0.34_{-0.10}^{+0.07}$	&	$0.42_{-0.13}^{+0.08}$	&	$0.35_{-0.14}^{+0.08}$	&	$0.17_{-0.10}^{+0.06}$	&	$0.18_{-0.04}^{+0.02}$	&	1.33	\\
E02 (*)	&	5.91	&	$0.01_{-0.01}^{+0.01}$	&	$0.56_{-0.03}^{+0.04}$	&	$8.09_{-2.19}^{+13.38}$	&	$24.64_{-5.33}^{+5.92}$	&	$0.23_{-0.05}^{+0.06}$	&	$0.33_{-0.07}^{+0.08}$	&	$0.31_{-0.08}^{+0.05}$	&	$0.41_{-0.12}^{+0.14}$	&	$0.14_{-0.02}^{+0.03}$	&	1.36	\\
E03 (*)	&	7.76	&	$0.01_{-0.01}^{+0.01}$	&	$0.63_{-0.07}^{+0.01}$	&	$12.87_{-6.34}^{+12.87}$	&	$25.70_{-6.00}^{+6.38}$	&	$0.18_{-0.03}^{+0.09}$	&	$0.24_{-0.07}^{+0.07}$	&	$0.26_{-0.13}^{+0.02}$	&	$0.12_{-0.08}^{+0.10}$	&	$0.14_{-0.04}^{+0.01}$	&	1.22	\\
E04 (*)	&	9.41	&	$0.01_{-0.01}^{+0.02}$	&	$0.62_{-0.04}^{+0.03}$	&	$5.71_{-1.11}^{+6.47}$	&	$23.14_{-5.10}^{+5.73}$	&	$0.22_{-0.05}^{+0.05}$	&	$0.29_{-0.07}^{+0.06}$	&	$0.22_{-0.08}^{+0.07}$	&	$0.21_{-0.09}^{+0.10}$	&	$0.15_{-0.02}^{+0.02}$	&	1.44	\\
E05 (*)	&	10.93	&	$0.01_{-0.01}^{+0.01}$	&	$0.65_{-0.05}^{+0.06}$	&	$5.15_{-0.67}^{+0.87}$	&	$21.50_{-2.67}^{+3.17}$	&	$0.22_{-0.05}^{+0.08}$	&	$0.25_{-0.08}^{+0.09}$	&	$0.25_{-0.08}^{+0.12}$	&	$0.30_{-0.11}^{+0.13}$	&	$0.16_{-0.03}^{+0.05}$	&	0.97	\\
E06 (*)	&	12.31	&	$0.01_{-0.01}^{+0.01}$	&	$0.72_{-0.08}^{+0.07}$	&	$7.30_{-1.59}^{+2.62}$	&	$23.91_{-3.01}^{+2.67}$	&	$0.18_{-0.09}^{+0.09}$	&	$0.22_{-0.10}^{+0.11}$	&	$0.13_{-0.08}^{+0.10}$	&	$0.08_{-0.08}^{+0.09}$	&	$0.12_{-0.03}^{+0.04}$	&	1.08	\\
E07	&	13.59	&	$0.01_{-0.01}^{+0.01}$	&	$0.64_{-0.05}^{+0.09}$	&	$5.67_{-2.57}^{+4.04}$	&	$25.60_{-4.66}^{+4.21}$	&	$0.22_{-0.06}^{+0.10}$	&	$0.25_{-0.08}^{+0.08}$	&	$0.14_{-0.03}^{+0.06}$	&	$0.23_{-0.08}^{+0.10}$	&	$0.11_{-0.02}^{+0.02}$	&	1.48	\\
E08	&	14.76	&	$0.01_{-0.01}^{+0.01}$	&	$0.72_{-0.12}^{+0.10}$	&	$2.81_{-1.21}^{+1.97}$	&	$19.32_{-3.73}^{+4.29}$	&	$0.19_{-0.05}^{+0.09}$	&	$0.22_{-0.04}^{+0.08}$	&	$0.17_{-0.06}^{+0.07}$	&	$0.25_{-0.09}^{+0.10}$	&	$0.15_{-0.03}^{+0.05}$	&	1.59	\\
E09	&	15.81	&	$0.01_{-0.01}^{+0.01}$	&	$0.65_{-0.06}^{+0.09}$	&	$3.49_{-1.52}^{+2.39}$	&	$22.46_{-4.32}^{+4.21}$	&	$0.17_{-0.04}^{+0.06}$	&	$0.21_{-0.06}^{+0.07}$	&	$0.17_{-0.06}^{+0.07}$	&	$0.26_{-0.09}^{+0.10}$	&	$0.11_{-0.02}^{+0.03}$	&	1.41	\\
E10	&	16.84	&	$0.01_{-0.01}^{+0.01}$	&	$0.59_{-0.04}^{+0.05}$	&	$5.32_{-2.08}^{+4.00}$	&	$25.89_{-3.96}^{+4.17}$	&	$0.20_{-0.05}^{+0.07}$	&	$0.23_{-0.05}^{+0.07}$	&	$0.20_{-0.06}^{+0.07}$	&	$0.21_{-0.08}^{+0.10}$	&	$0.11_{-0.02}^{+0.03}$	&	1.14	\\
E11	&	17.86	&	$0.01_{-0.01}^{+0.01}$	&	$0.63_{-0.06}^{+0.11}$	&	$2.72_{-1.20}^{+1.59}$	&	$22.77_{-5.04}^{+4.13}$	&	$0.13_{-0.03}^{+0.04}$	&	$0.26_{-0.06}^{+0.06}$	&	$0.17_{-0.06}^{+0.06}$	&	$0.31_{-0.10}^{+0.11}$	&	$0.12_{-0.02}^{+0.04}$	&	1.47	\\
E12	&	18.84	&	$0.01_{-0.01}^{+0.01}$	&	$0.60_{-0.05}^{+0.08}$	&	$2.82_{-1.08}^{+1.60}$	&	$22.35_{-4.27}^{+4.12}$	&	$0.14_{-0.03}^{+0.04}$	&	$0.25_{-0.05}^{+0.06}$	&	$0.20_{-0.06}^{+0.07}$	&	$0.30_{-0.10}^{+0.12}$	&	$0.14_{-0.02}^{+0.02}$	&	1.56	\\
E13	&	19.77	&	$0.01_{-0.01}^{+0.01}$	&	$0.58_{-0.08}^{+0.06}$	&	$3.45_{-1.30}^{+2.24}$	&	$26.35_{-4.44}^{+4.70}$	&	$0.13_{-0.03}^{+0.04}$	&	$0.25_{-0.05}^{+0.05}$	&	$0.22_{-0.06}^{+0.07}$	&	$0.22_{-0.09}^{+0.10}$	&	$0.12_{-0.02}^{+0.03}$	&	1.22	\\
E14	&	20.65	&	$0.01_{-0.01}^{+0.01}$	&	$0.61_{-0.05}^{+0.07}$	&	$2.92_{-1.09}^{+1.86}$	&	$24.73_{-4.50}^{+4.72}$	&	$0.11_{-0.02}^{+0.03}$	&	$0.22_{-0.05}^{+0.05}$	&	$0.15_{-0.06}^{+0.06}$	&	$0.29_{-0.10}^{+0.11}$	&	$0.11_{-0.02}^{+0.03}$	&	1.14	\\
E15	&	21.46	&	$0.01_{-0.01}^{+0.01}$	&	$0.65_{-0.06}^{+0.11}$	&	$2.61_{-1.21}^{+1.69}$	&	$21.58_{-4.84}^{+4.25}$	&	$0.14_{-0.03}^{+0.04}$	&	$0.21_{-0.06}^{+0.06}$	&	$0.10_{-0.05}^{+0.06}$	&	$0.22_{-0.09}^{+0.09}$	&	$0.13_{-0.02}^{+0.04}$	&	1.33	\\
E16	&	22.25	&	$0.01_{-0.01}^{+0.01}$	&	$0.74_{-0.10}^{+0.13}$	&	$1.66_{-0.70}^{+1.27}$	&	$20.88_{-4.62}^{+5.22}$	&	$0.12_{-0.02}^{+0.03}$	&	$0.15_{-0.04}^{+0.05}$	&	$0.11_{-0.05}^{+0.06}$	&	$0.20_{-0.07}^{+0.09}$	&	$0.14_{-0.03}^{+0.04}$	&	1.38	\\
E17	&	23.04	&	$0.01_{-0.01}^{+0.01}$	&	$0.71_{-0.09}^{+0.10}$	&	$1.77_{-0.69}^{+1.40}$	&	$20.63_{-4.07}^{+5.57}$	&	$0.10_{-0.02}^{+0.03}$	&	$0.17_{-0.04}^{+0.04}$	&	$0.11_{-0.05}^{+0.05}$	&	$0.19_{-0.08}^{+0.09}$	&	$0.12_{-0.03}^{+0.03}$	&	1.42	\\
E18	&	23.81	&	$0.01_{-0.01}^{+0.01}$	&	$0.72_{-0.08}^{+0.13}$	&	$2.01_{-0.86}^{+1.25}$	&	$18.20_{-2.05}^{+3.79}$	&	$0.16_{-0.03}^{+0.05}$	&	$0.25_{-0.06}^{+0.06}$	&	$0.17_{-0.06}^{+0.07}$	&	$0.20_{-0.08}^{+0.10}$	&	$0.15_{-0.03}^{+0.05}$	&	1.49	\\
E19	&	24.64	&	$0.01_{-0.01}^{+0.01}$	&	$0.79_{-0.09}^{+0.20}$	&	$1.18_{-0.50}^{+0.57}$	&	$16.17_{-4.13}^{+3.47}$	&	$0.13_{-0.02}^{+0.03}$	&	$0.29_{-0.06}^{+0.06}$	&	$0.19_{-0.06}^{+0.07}$	&	$0.26_{-0.10}^{+0.10}$	&	$0.15_{-0.03}^{+0.05}$	&	1.37	\\
E20	&	25.51	&	$0.01_{-0.01}^{+0.01}$	&	$0.76_{-0.09}^{+0.17}$	&	$1.61_{-0.72}^{+0.94}$	&	$17.71_{-4.43}^{+3.80}$	&	$0.15_{-0.03}^{+0.04}$	&	$0.28_{-0.05}^{+0.06}$	&	$0.21_{-0.06}^{+0.07}$	&	$0.17_{-0.08}^{+0.09}$	&	$0.12_{-0.03}^{+0.04}$	&	1.09	\\
E21	&	26.37	&	$0.01_{-0.01}^{+0.01}$	&	$0.62_{-0.06}^{+0.12}$	&	$3.39_{-1.56}^{+2.06}$	&	$23.00_{-5.67}^{+3.56}$	&	$0.15_{-0.03}^{+0.05}$	&	$0.34_{-0.06}^{+0.07}$	&	$0.19_{-0.03}^{+0.07}$	&	$0.27_{-0.10}^{+0.10}$	&	$0.08_{-0.02}^{+0.03}$	&	1.59	\\
E22	&	27.21	&	$0.01_{-0.01}^{+0.01}$	&	$0.70_{-0.12}^{+0.11}$	&	$1.41_{-0.51}^{+1.36}$	&	$18.50_{-1.92}^{+6.65}$	&	$0.13_{-0.02}^{+0.03}$	&	$0.27_{-0.05}^{+0.05}$	&	$0.15_{-0.06}^{+0.06}$	&	$0.31_{-0.10}^{+0.14}$	&	$0.11_{-0.05}^{+0.03}$	&	1.59	\\
E23	&	28.08	&	$0.01_{-0.01}^{+0.01}$	&	$0.82_{-0.13}^{+0.23}$	&	$0.73_{-0.33}^{+0.41}$	&	$14.59_{-4.06}^{+3.84}$	&	$0.12_{-0.02}^{+0.02}$	&	$0.27_{-0.05}^{+0.06}$	&	$0.18_{-0.06}^{+0.08}$	&	$0.28_{-0.11}^{+0.11}$	&	$0.15_{-0.03}^{+0.04}$	&	1.33	\\
E24	&	29.09	&	$0.01_{-0.01}^{+0.01}$	&	$0.60_{-0.07}^{+0.11}$	&	$2.45_{-1.16}^{+1.93}$	&	$24.03_{-5.92}^{+5.82}$	&	$0.14_{-0.03}^{+0.04}$	&	$0.22_{-0.04}^{+0.05}$	&	$0.13_{-0.05}^{+0.06}$	&	$0.19_{-0.09}^{+0.10}$	&	$0.08_{-0.02}^{+0.03}$	&	1.36	\\
\hline
F01 (*)	&	3.46	&	$0.05_{-0.03}^{+0.03}$	&	$1.31_{-0.17}^{+0.15}$	&	$5.56_{-1.22}^{+1.36}$	&	$21.52_{-2.23}^{+2.75}$	&	$1.37_{-0.46}^{+0.93}$	&	$1.05_{-0.62}^{+1.14}$	&	$1.34_{-0.51}^{+1.14}$	&	$0.55_{-0.28}^{+0.58}$	&	$0.71_{-0.31}^{+0.59}$	&	1.47	\\
F02 (*)	&	6.44	&	$0.01_{-0.01}^{+0.11}$	&	$0.83_{-0.53}^{+2.01}$	&	$3.71_{-1.85}^{+10.64}$	&	$15.51_{-0.98}^{+1.27}$	&	$0.66_{-1.23}^{+3.55}$	&	$0.81_{-4.42}^{+0.81}$	&	$0.50_{-0.85}^{+3.57}$	&	$0.44_{-0.15}^{+1.83}$	&	$0.19_{-0.19}^{+0.09}$	&	1.28	\\
F03 (*)	&	8.66	&	$0.08_{-0.01}^{+0.05}$	&	$1.29_{-0.34}^{+4.32}$	&	$4.01_{-1.66}^{+13.32}$	&	$16.55_{-1.12}^{+1.68}$	&	$1.04_{-0.10}^{+3.08}$	&	$1.36_{-1.87}^{+1.36}$	&	$0.58_{-0.25}^{+2.98}$	&	$0.37_{-0.08}^{+1.18}$	&	$0.23_{-0.23}^{+0.02}$	&	1.25	\\
F04 (*)	&	10.53	&	$0.01_{-0.01}^{+0.05}$	&	$1.01_{-0.13}^{+0.12}$	&	$5.48_{-1.05}^{+1.74}$	&	$12.05_{-1.97}^{+1.95}$	&	$1.15_{-0.35}^{+0.42}$	&	$0.87_{-0.60}^{+0.91}$	&	$0.67_{-0.29}^{+0.59}$	&	$0.33_{-0.18}^{+0.33}$	&	$0.40_{-0.16}^{+0.30}$	&	1.36	\\
F05 (*)	&	11.92	&	$0.02_{-0.02}^{+0.05}$	&	$1.14_{-0.15}^{+0.17}$	&	$3.79_{-0.79}^{+1.30}$	&	$12.22_{-1.82}^{+1.74}$	&	$1.00_{-0.34}^{+0.57}$	&	$0.83_{-0.53}^{+1.09}$	&	$0.95_{-0.37}^{+0.83}$	&	$0.33_{-0.19}^{+0.33}$	&	$0.47_{-0.20}^{+0.34}$	&	0.94	\\
F06 (*)	&	13.07	&	$0.01_{-0.01}^{+0.10}$	&	$0.91_{-0.10}^{+0.58}$	&	$2.79_{-0.23}^{+0.27}$	&	$14.23_{-0.51}^{+13.12}$	&	$0.33_{-0.08}^{+0.29}$	&	$0.43_{-0.11}^{+0.59}$	&	$0.29_{-0.09}^{+0.47}$	&	$0.14_{-0.09}^{+0.08}$	&	$0.18_{-0.06}^{+0.07}$	&	1.26	\\
F07	&	14.18	&	$0.01_{-0.01}^{+0.01}$	&	$0.70_{-0.08}^{+0.08}$	&	$8.32_{-3.58}^{+6.60}$	&	$20.27_{-3.13}^{+4.14}$	&	$0.44_{-0.14}^{+0.20}$	&	$0.71_{-0.18}^{+0.25}$	&	$0.26_{-0.09}^{+0.10}$	&	$0.26_{-0.09}^{+0.12}$	&	$0.09_{-0.02}^{+0.03}$	&	1.31	\\
F08	&	15.32	&	$0.01_{-0.01}^{+0.01}$	&	$0.74_{-0.03}^{+0.09}$	&	$8.70_{-4.20}^{+6.52}$	&	$20.77_{-3.38}^{+3.23}$	&	$0.42_{-0.14}^{+0.20}$	&	$0.47_{-0.17}^{+0.19}$	&	$0.29_{-0.09}^{+0.10}$	&	$0.22_{-0.09}^{+0.10}$	&	$0.09_{-0.02}^{+0.04}$	&	1.30	\\
F09	&	16.47	&	$0.01_{-0.01}^{+0.01}$	&	$0.71_{-0.08}^{+0.07}$	&	$7.91_{-3.44}^{+5.49}$	&	$21.04_{-3.09}^{+3.59}$	&	$0.37_{-0.12}^{+0.18}$	&	$0.46_{-0.14}^{+0.18}$	&	$0.24_{-0.08}^{+0.09}$	&	$0.19_{-0.08}^{+0.09}$	&	$0.10_{-0.02}^{+0.03}$	&	0.93	\\
F10	&	17.58	&	$0.01_{-0.01}^{+0.01}$	&	$0.69_{-0.07}^{+0.08}$	&	$5.44_{-2.30}^{+4.29}$	&	$22.32_{-3.38}^{+4.24}$	&	$0.26_{-0.08}^{+0.12}$	&	$0.34_{-0.09}^{+0.13}$	&	$0.20_{-0.07}^{+0.08}$	&	$0.19_{-0.08}^{+0.10}$	&	$0.11_{-0.02}^{+0.03}$	&	1.30	\\
F11	&	18.65	&	$0.01_{-0.01}^{+0.01}$	&	$0.62_{-0.05}^{+0.07}$	&	$4.61_{-1.80}^{+3.21}$	&	$25.38_{-4.17}^{+4.55}$	&	$0.16_{-0.04}^{+0.06}$	&	$0.29_{-0.06}^{+0.08}$	&	$0.15_{-0.06}^{+0.06}$	&	$0.27_{-0.09}^{+0.11}$	&	$0.10_{-0.02}^{+0.02}$	&	1.26	\\
F12	&	19.65	&	$0.01_{-0.01}^{+0.01}$	&	$0.60_{-0.05}^{+0.07}$	&	$4.82_{-1.99}^{+3.78}$	&	$25.58_{-4.46}^{+4.53}$	&	$0.18_{-0.04}^{+0.06}$	&	$0.26_{-0.06}^{+0.07}$	&	$0.14_{-0.05}^{+0.06}$	&	$0.24_{-0.09}^{+0.10}$	&	$0.09_{-0.02}^{+0.02}$	&	1.33	\\
F13	&	20.68	&	$0.01_{-0.01}^{+0.01}$	&	$0.62_{-0.06}^{+0.12}$	&	$3.04_{-1.37}^{+1.90}$	&	$23.19_{-5.41}^{+4.58}$	&	$0.17_{-0.03}^{+0.05}$	&	$0.32_{-0.06}^{+0.06}$	&	$0.17_{-0.06}^{+0.07}$	&	$0.38_{-0.11}^{+0.12}$	&	$0.08_{-0.02}^{+0.03}$	&	1.59	\\
F14	&	21.86	&	$0.01_{-0.01}^{+0.01}$	&	$0.71_{-0.10}^{+0.13}$	&	$2.30_{-1.08}^{+2.17}$	&	$19.73_{-4.42}^{+5.69}$	&	$0.18_{-0.04}^{+0.06}$	&	$0.24_{-0.05}^{+0.06}$	&	$0.13_{-0.05}^{+0.06}$	&	$0.17_{-0.08}^{+0.09}$	&	$0.09_{-0.03}^{+0.03}$	&	1.59	\\
F15	&	23.20	&	$0.01_{-0.01}^{+0.01}$	&	$0.62_{-0.06}^{+0.11}$	&	$2.58_{-1.10}^{+1.61}$	&	$24.97_{-5.60}^{+5.09}$	&	$0.16_{-0.03}^{+0.04}$	&	$0.28_{-0.04}^{+0.05}$	&	$0.13_{-0.05}^{+0.06}$	&	$0.25_{-0.09}^{+0.10}$	&	$0.07_{-0.01}^{+0.02}$	&	1.24	\\
F16	&	24.82	&	$0.01_{-0.01}^{+0.01}$	&	$0.64_{-0.06}^{+0.13}$	&	$2.82_{-1.29}^{+1.71}$	&	$24.16_{-5.87}^{+4.85}$	&	$0.15_{-0.03}^{+0.04}$	&	$0.28_{-0.05}^{+0.05}$	&	$0.17_{-0.06}^{+0.06}$	&	$0.18_{-0.08}^{+0.09}$	&	$0.08_{-0.02}^{+0.03}$	&	1.24	\\
F17	&	26.62	&	$0.01_{-0.01}^{+0.01}$	&	$0.63_{-0.06}^{+0.13}$	&	$2.77_{-1.26}^{+1.71}$	&	$23.61_{-5.60}^{+4.81}$	&	$0.15_{-0.03}^{+0.04}$	&	$0.27_{-0.05}^{+0.05}$	&	$0.12_{-0.05}^{+0.06}$	&	$0.22_{-0.08}^{+0.10}$	&	$0.09_{-0.02}^{+0.03}$	&	1.33	\\
F18	&	28.37	&	$0.01_{-0.01}^{+0.01}$	&	$0.69_{-0.10}^{+0.11}$	&	$1.56_{-0.61}^{+1.30}$	&	$19.26_{-4.13}^{+5.88}$	&	$0.14_{-0.02}^{+0.03}$	&	$0.27_{-0.04}^{+0.05}$	&	$0.19_{-0.06}^{+0.07}$	&	$0.17_{-0.08}^{+0.10}$	&	$0.10_{-0.03}^{+0.02}$	&	1.59	\\
F19	&	30.07	&	$0.01_{-0.01}^{+0.01}$	&	$0.70_{-0.12}^{+0.12}$	&	$1.03_{-0.42}^{+1.04}$	&	$18.36_{-4.20}^{+7.21}$	&	$0.12_{-0.02}^{+0.02}$	&	$0.26_{-0.04}^{+0.05}$	&	$0.17_{-0.06}^{+0.07}$	&	$0.25_{-0.09}^{+0.11}$	&	$0.09_{-0.03}^{+0.03}$	&	1.26	\\
F20	&	31.90	&	$0.01_{-0.01}^{+0.01}$	&	$0.56_{-0.07}^{+0.08}$	&	$2.95_{-1.27}^{+2.94}$	&	$27.15_{-5.67}^{+9.15}$	&	$0.16_{-0.03}^{+0.04}$	&	$0.28_{-0.04}^{+0.05}$	&	$0.11_{-0.05}^{+0.05}$	&	$0.23_{-0.10}^{+0.11}$	&	$0.05_{-0.01}^{+0.02}$	&	1.20	\\
\hline
    \end{tabular}%
\\
Note: Abundances are with respect to solar \citep{Anders89}. Uncertainties are at the 90\% confidence level. The Galactic column $N_{{\rm H,Gal}}$ is fixed at $1.72\times10^{21}$ cm$^{-2}$ \citep{HI4PI16}. \\
(*) Double shock model parameters. For these regions the best-fit parameters of the ejecta component are presented.
\end{table*}

\subsubsection{Inner and Outer Faint Structures on N63A}
 
We perform spectral analysis of selected 12 X-ray faint regions from the inner and outer parts of the remnant (see Figure 2c) to investigate the detailed features these ``crescent-like'' and the ``hole-like'' structures. These regions were modelled with a single component plane shock model by varying $kT$, $n_{\rm e}t$, normalisation and elemental abundances (O, Ne, Mg, Si, and Fe) and statistically acceptable model fits were obtained ($\chi^2_{\nu}<1.6$). In Figure 6, we show some example spectra extracted from the regions marked in Figure 2c with best-fit models and residuals. The best-fit model parameters for these regions are listed in Table \ref{tab:five_inner_outer}. The spectral parameters for the outer diffuse and faint regions (Ear 1 - Ear 3 and Tail 1 - Tail 4) are compatible with median ISM values within statistical uncertainties. It is seen that $kT$ and $n_{\rm e}t$ for the inner faint regions (I1-I5) of the remnant are generally higher than the median ISM values, and the elemental abundances are consistent with the ISM values within the error limits.

\begin{table*}
\setlength{\tabcolsep}{2.5pt}
\renewcommand{\arraystretch}{1.5}
\small
  \centering
  \caption{Summary of spectral model fits to seven outer and five inner regions of N63A.}
    \begin{tabular}{ccccccccccc}
\hline
Region    &    $n_{\rm H}$             &   $kT$  & $n_{\rm e}t$                & $EM$                 & O    & Ne  & Mg  & Si & Fe & $\chi^{2}_{\nu}$\\
    &($10^{22}$cm$^{-2}$)  &  (keV)  & ($10^{11}$cm$^{-3}$s) & ($10^{57}$cm$^{-3}$) &      &     &     &    &    &                 \\
\hline
Ear 1  & $0.04_{-0.03}^{+0.08}$ & $0.57_{-0.09}^{+0.10}$ & $ 0.85_{-0.34}^{+0.73}$ & $ 14.90_{-4.68}^{+4.17}$ & $0.19_{-0.03}^{+0.03}$ & $0.39_{-0.07}^{+0.08}$ & $0.34_{-0.10}^{+0.12}$ & $0.62_{-0.22}^{+0.28}$ & $0.21_{-0.05}^{+0.07}$ & 1.55 \\
Ear 2  & $0.01_{-0.01}^{+0.01}$ & $0.57_{-0.09}^{+0.10}$ & $ 0.65_{-0.26}^{+0.59}$ & $ 14.22_{-3.46}^{+5.96}$ & $0.18_{-0.03}^{+0.02}$ & $0.27_{-0.05}^{+0.06}$ & $0.24_{-0.08}^{+0.10}$ & $0.25_{-0.15}^{+0.18}$ & $0.25_{-0.07}^{+0.09}$ & 1.30 \\
Ear 3  & $0.01_{-0.01}^{+0.07}$ & $0.65_{-0.09}^{+0.14}$ & $ 0.76_{-0.34}^{+0.57}$ & $ 24.00_{-6.65}^{+7.79}$ & $0.16_{-0.02}^{+0.02}$ & $0.26_{-0.04}^{+0.05}$ & $0.20_{-0.06}^{+0.07}$ & $0.26_{-0.10}^{+0.11}$ & $0.18_{-0.04}^{+0.07}$ & 1.58 \\

Tail 1 & $0.01_{-0.01}^{+0.03}$ & $0.54_{-0.11}^{+0.17}$ & $ 1.29_{-0.76}^{+2.63}$ & $ 27.33_{-10.08}^{+15.48}$ & $0.16_{-0.02}^{+0.03}$ & $0.27_{-0.04}^{+0.07}$ & $0.17_{-0.05}^{+0.06}$ & $0.34_{-0.12}^{+0.13}$ & $0.06_{-0.02}^{+0.04}$ & 1.59 \\

Tail 2 & $0.01_{-0.01}^{+0.01}$ & $0.56_{-0.08}^{+0.54}$ & $ 2.20_{-1.88}^{+3.80}$ & $ 28.94_{-17.77}^{+5.59}$ & $0.18_{-0.02}^{+0.03}$ & $0.26_{-0.01}^{+0.11}$ & $0.05_{-0.03}^{+0.07}$ & $0.17_{-0.15}^{+0.03}$ & $0.01_{-0.01}^{+0.05}$ & 1.57 \\

Tail 3 & $0.01_{-0.01}^{+0.01}$ & $0.77_{-0.07}^{+0.12}$ & $ 0.98_{-0.34}^{+0.71}$ & $ 24.40_{-4.99}^{+7.62}$ & $0.13_{-0.01}^{+0.01}$ & $0.32_{-0.03}^{+0.04}$ & $0.17_{-0.02}^{+0.04}$ & $0.22_{-0.06}^{+0.07}$ & $0.09_{-0.02}^{+0.01}$ & 1.51 \\

Tail 4 & $0.01_{-0.01}^{+0.02}$ & $0.68_{-0.11}^{+0.10}$ & $ 0.53_{-0.16}^{+0.39}$ & $ 17.90_{-2.85}^{+6.78}$ & $0.17_{-0.02}^{+0.02}$ & $0.26_{-0.04}^{+0.04}$ & $0.12_{-0.06}^{+0.06}$ & $0.19_{-0.11}^{+0.12}$ & $0.08_{-0.03}^{+0.02}$ & 1.49 \\

\hline
I1 & $0.16_{-0.05}^{+0.06}$ & $0.71_{-0.04}^{+0.05}$ & $ 13.06_{~-7.0}^{+17.2}$  & $ 20.14_{-4.38}^{+5.01}$ & $0.42_{-0.19}^{+0.32}$ & $0.37_{-0.18}^{+0.31}$ & $0.48_{-0.16}^{+0.25}$ & $0.39_{-0.12}^{+0.17}$ & $0.16_{-0.04}^{+0.05}$ & 1.07 \\
I2 & $0.23_{-0.07}^{+0.10}$ & $0.73_{-0.05}^{+0.05}$ & $ 48.62_{-4.92}^{+50.0}$ & $ 11.75_{-3.60}^{+4.57}$ & $0.84_{-0.29}^{+0.97}$ & $1.34_{-0.85}^{+1.17}$ & $1.06_{-0.41}^{+0.57}$ & $0.34_{-0.17}^{+0.26}$ & $0.15_{-0.05}^{+0.08}$ & 1.14 \\
I3 & $0.30_{-0.07}^{+0.08}$ & $0.79_{-0.06}^{+0.08}$ & $ 28.64_{-20.2}^{+66.2}$ & $ 21.30_{-4.08}^{+4.56}$ & $0.40_{-0.21}^{+0.44}$ & $0.36_{-0.24}^{+0.35}$ & $0.37_{-0.15}^{+0.22}$ & $0.40_{-0.12}^{+0.16}$ & $0.09_{-0.02}^{+0.03}$ & 1.18 \\
I4 & $0.34_{-0.06}^{+0.07}$ & $0.75_{-0.04}^{+0.03}$ & $ 36.98_{-11.8}^{+23.4}$ & $ 41.96_{-6.23}^{+8.01}$ & $0.43_{-0.20}^{+0.21}$ & $0.84_{-0.27}^{+0.28}$ & $0.43_{-0.13}^{+0.16}$ & $0.33_{-0.09}^{+0.10}$ & $0.05_{-0.01}^{+0.02}$ & 0.86 \\
I5 & $0.25_{-0.08}^{+0.09}$ & $0.77_{-0.07}^{+0.19}$ & $ 32.68_{-30.1}^{+43.5}$ & $ 15.72_{-3.12}^{+4.00}$ & $0.19_{-0.14}^{+0.26}$ & $0.15_{-0.15}^{+0.17}$ & $0.35_{-0.16}^{+0.24}$ & $0.20_{-0.10}^{+0.13}$ & $0.05_{-0.02}^{+0.04}$ & 0.93 \\
\hline
    \end{tabular}%
    \label{tab:five_inner_outer}
\\
Note: Abundances are with respect to solar \citep{Anders89}. Uncertainties are at the 90\% confidence level. The Galactic column $N_{{\rm H,Gal}}$ is fixed at $1.72\times10^{21}$ cm$^{-2}$ \citep{HI4PI16}. 
\end{table*}%

\begin{figure*}
\begin{center}
\includegraphics[width=\textwidth]{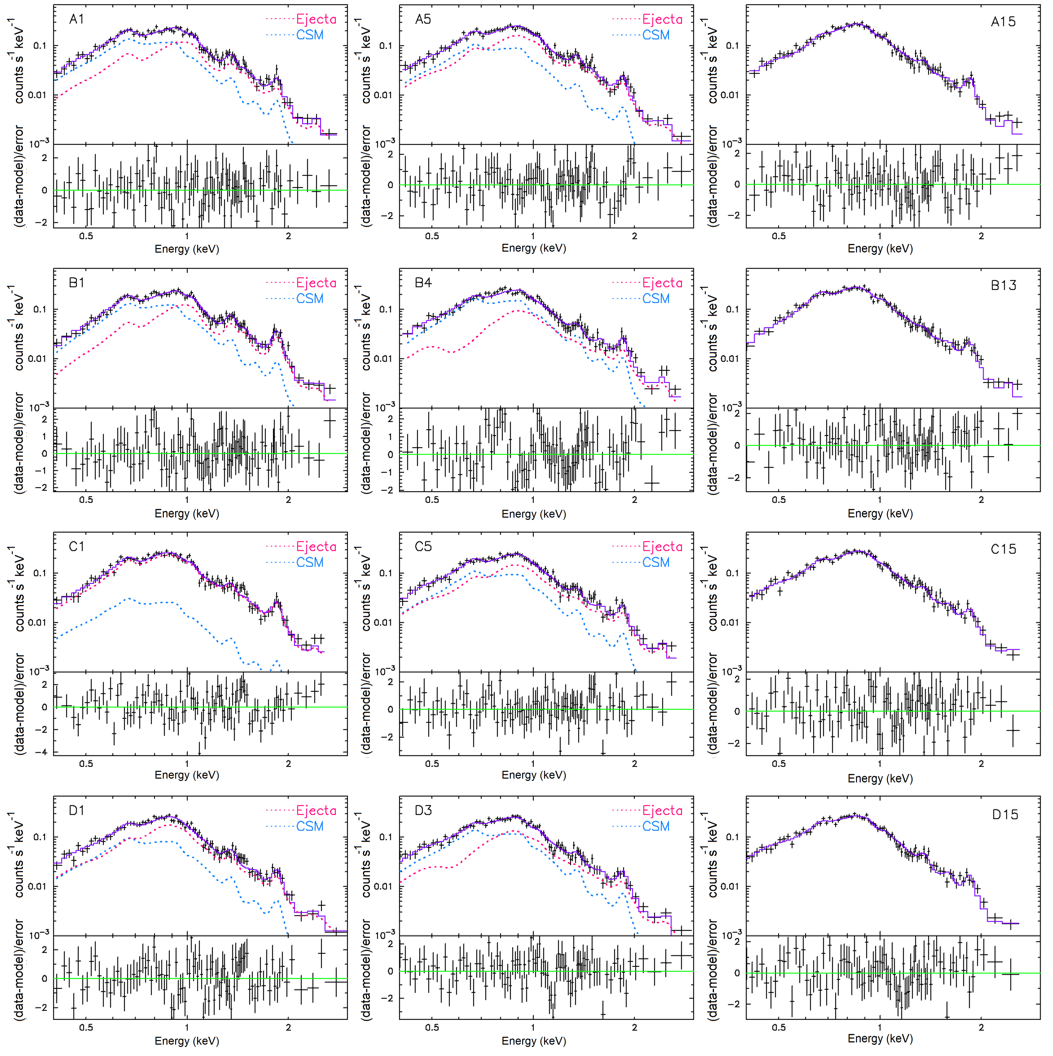}
\caption{
A set of sample best-fit models and residuals of X-ray spectra from selected regions shown in Figure 2b.}
\end{center}
\end{figure*}

\begin{figure*}
\begin{center}
\includegraphics[width=\textwidth]{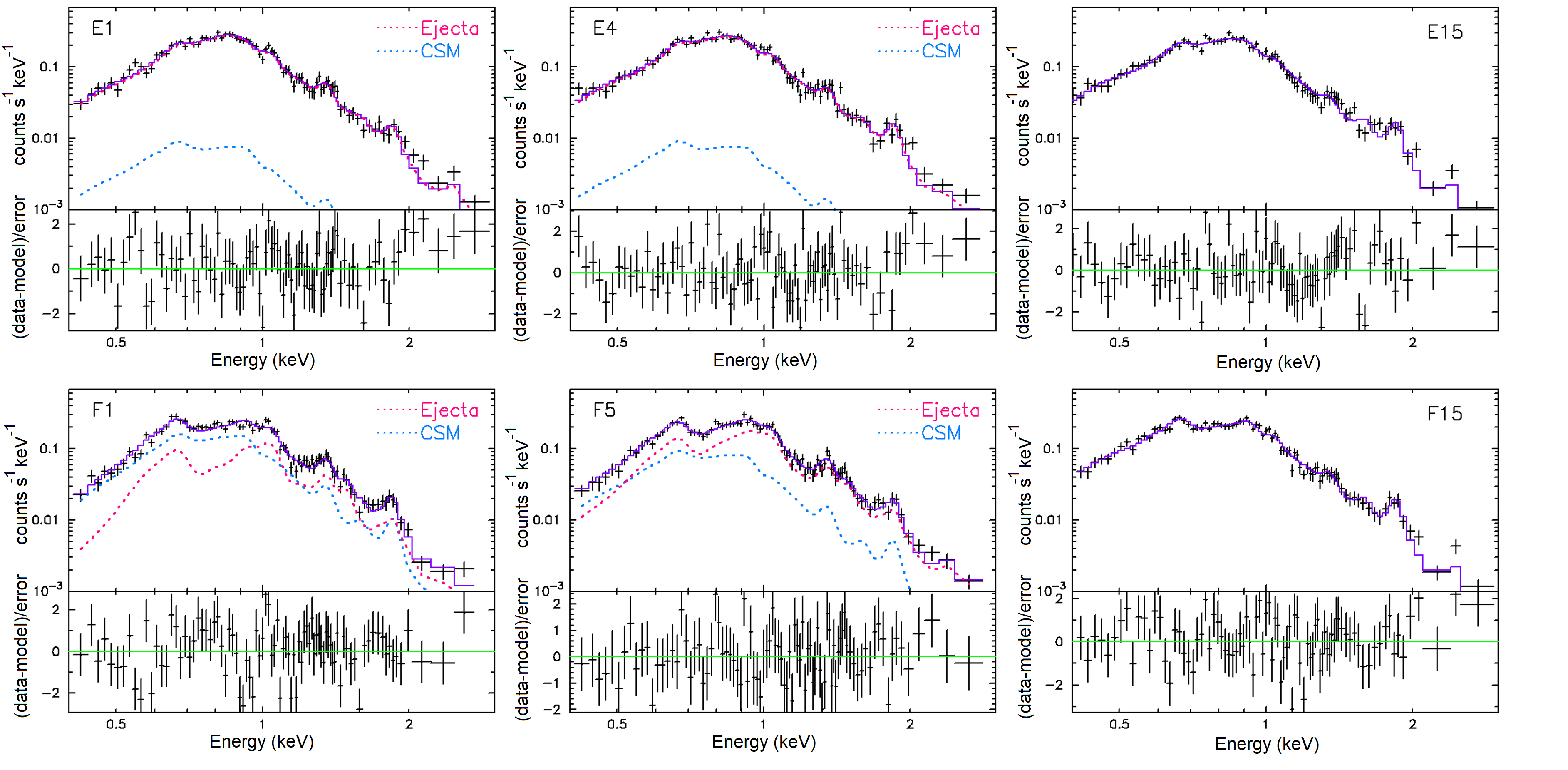}
\caption{
 A set of sample best-fit models and residuals of X-ray spectra from selected regions shown in Figure 2b.}
\end{center}
\end{figure*}

\begin{figure*}
\begin{center}
\includegraphics[width=\textwidth]{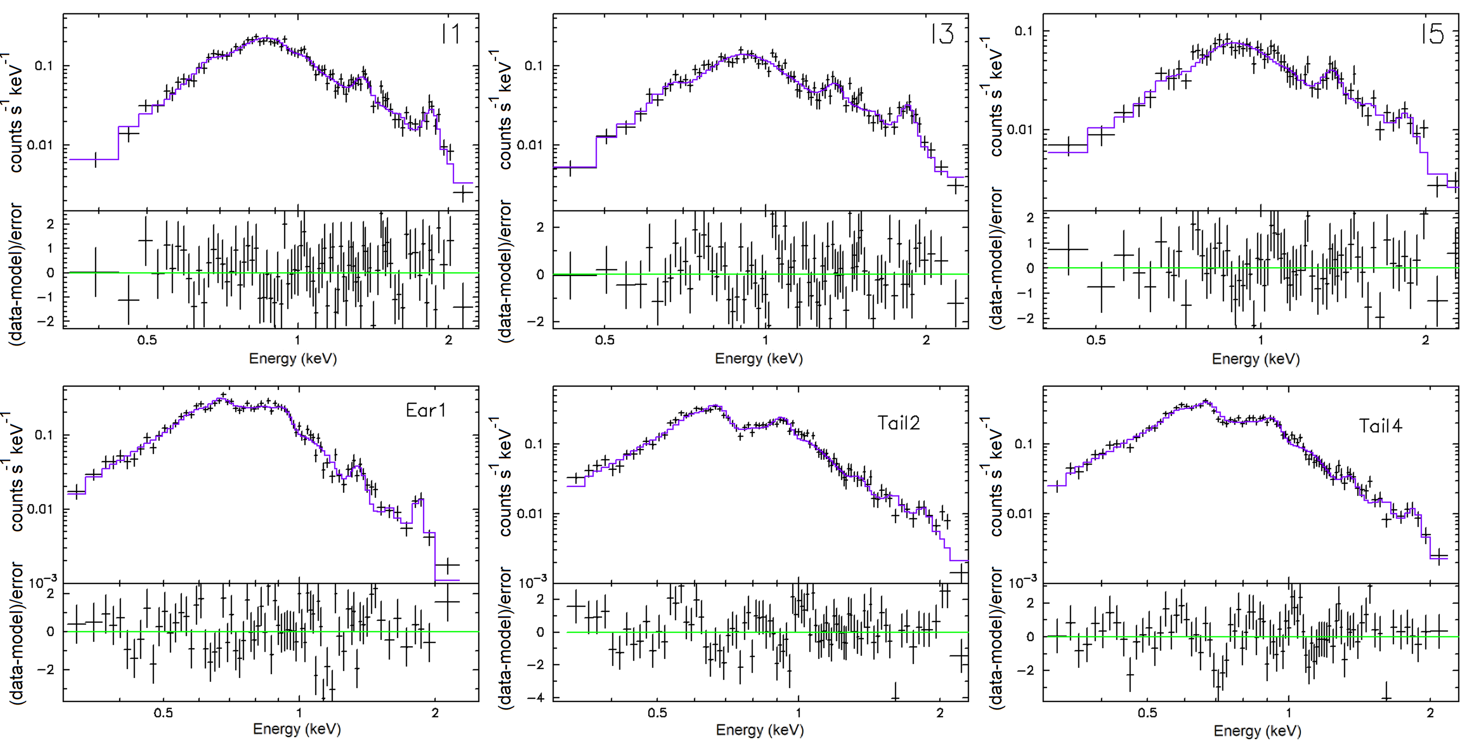}
\caption{
 A set of sample best-fit models and residuals of X-ray spectra from selected regions are shown in Figure 2c.}
\end{center}
\end{figure*}

\begin{figure*}
\begin{center}
\includegraphics[width=\textwidth]{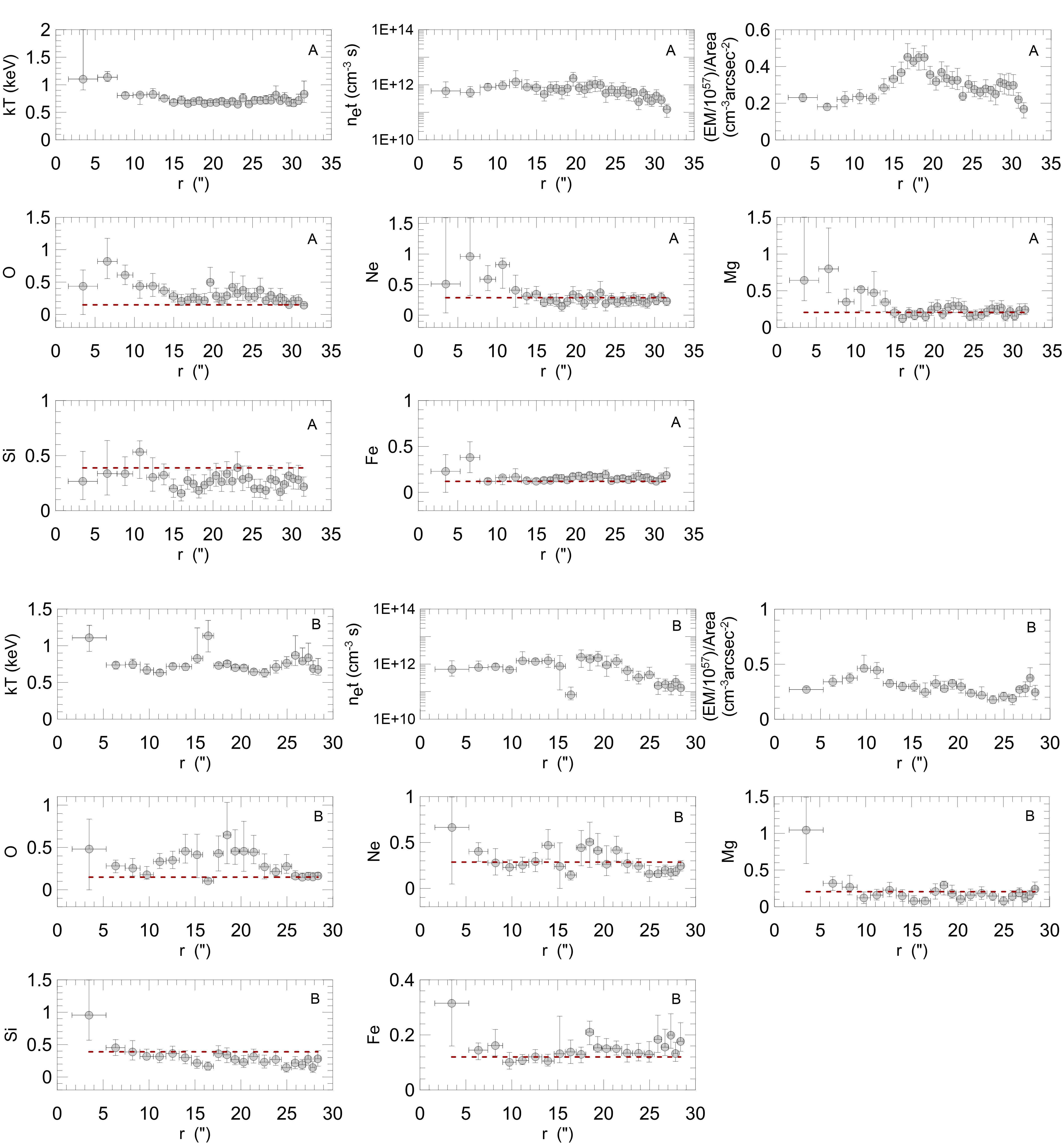}
\caption{Best-fit spectral parameters with error bars along the radius of N63A. The parameters for the ejecta component are electron temperature ($kT$), ionization timescale ($n_et$), emission measure ($EM$), and O, Ne, Mg, Si, and Fe abundances in the A and B directions of N63A. In $EM$ plots the broadband surface brightness profile is overlaid with a green line. The red dashed lines in the abundance panels are the mean shell abundances for each element. Abundances are with respect to solar \citep{Anders89}.}
\end{center}
\end{figure*}

\begin{figure*}
\begin{center}
\includegraphics[width=\textwidth]{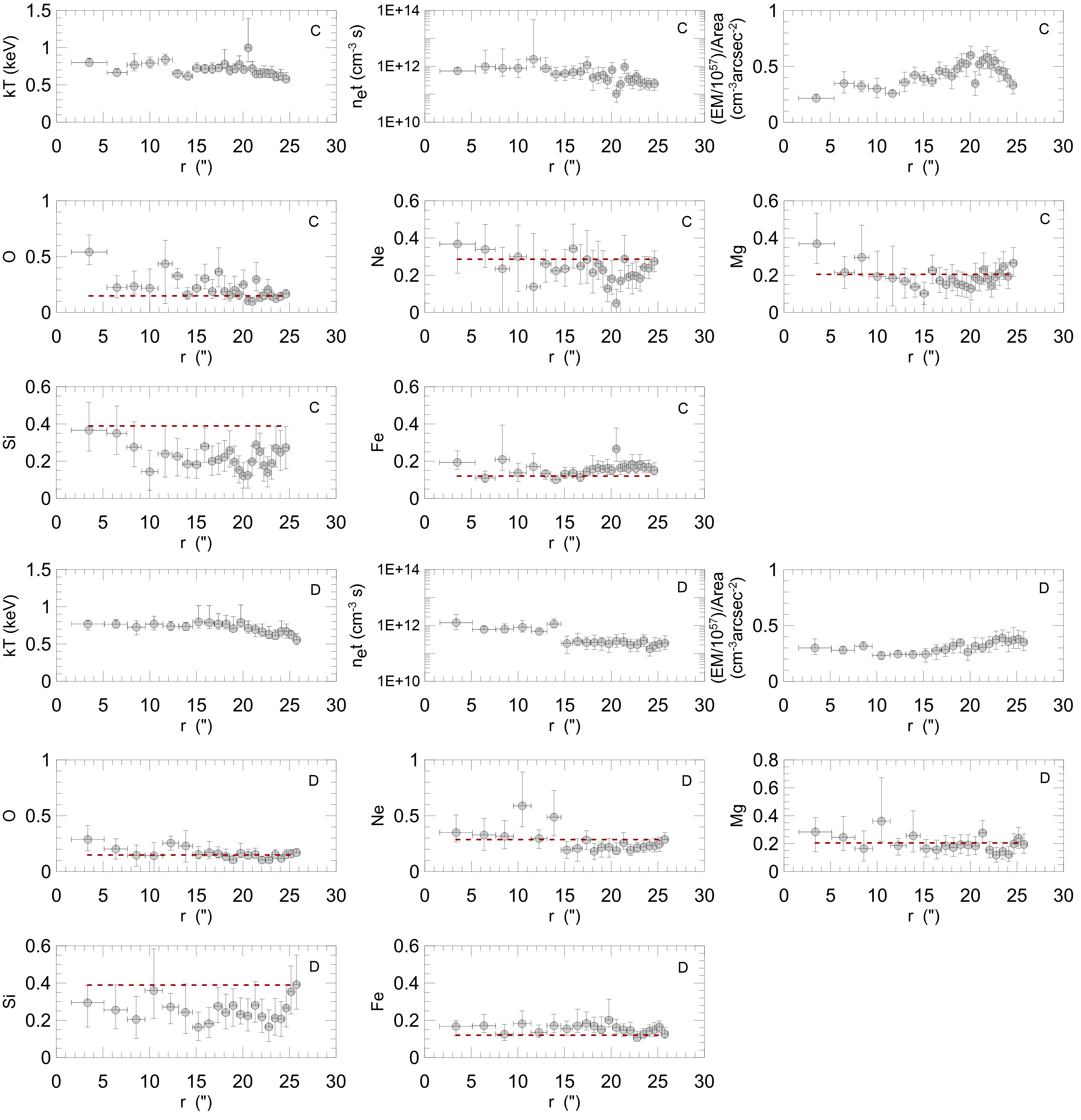}
\caption{Same as spectral parameters in Figure 7 but for the C and D directions of N63A.}
\end{center}
\end{figure*}

\begin{figure*}
\begin{center}
\includegraphics[width=\textwidth]{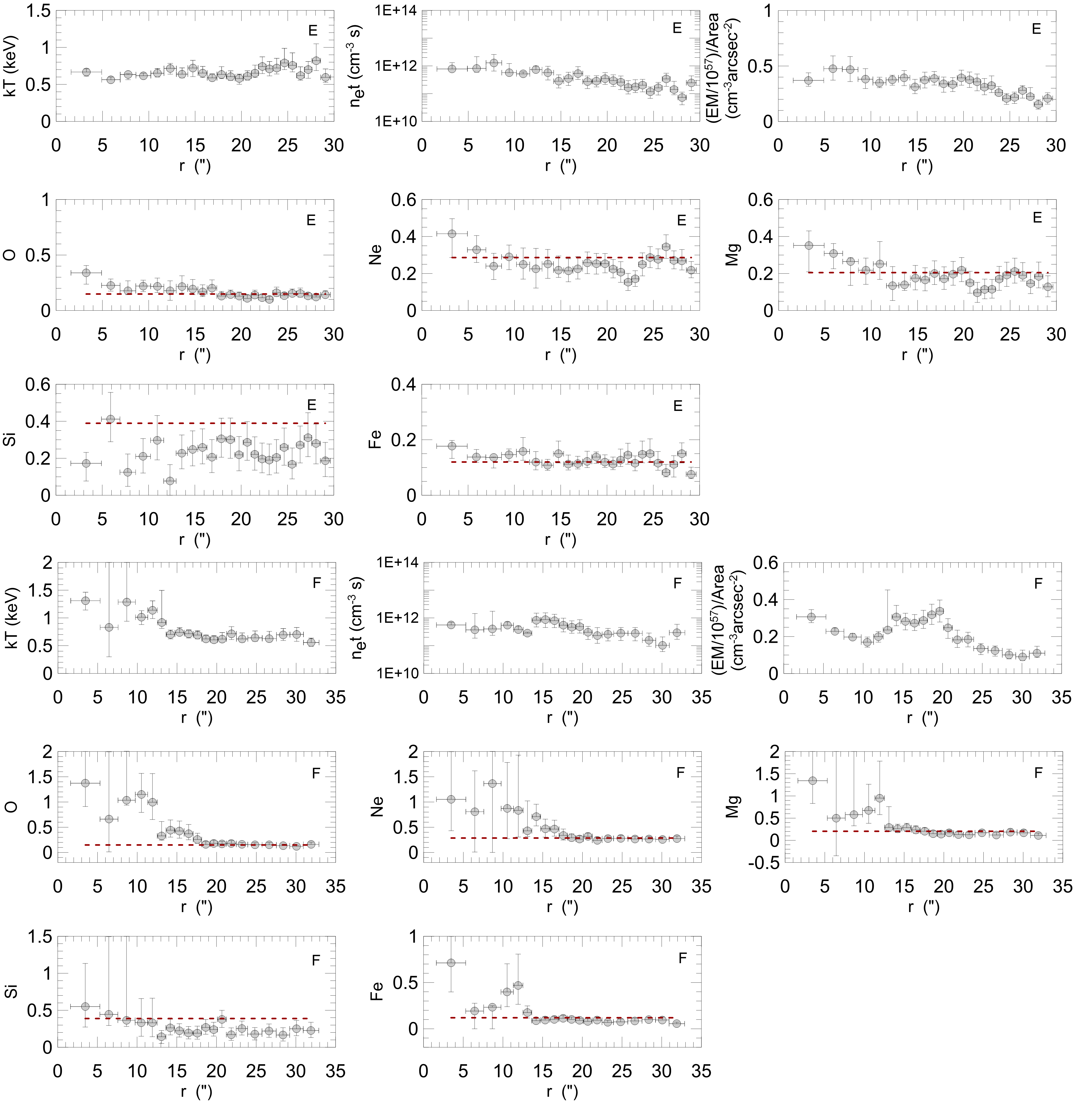}
\caption{Same as spectral parameters in Figure 7 but for the E and F directions of N63A.}
\end{center}
\end{figure*}

\section{Discussion}
\subsection{Morphology of the Remnant}
The two main morphological features of the N63A are detached by imaging; the ``crescent-like'' and the ``hole-like'' structures. \citet{Warren03} suggested that the origin of ``crescent-like'' structures located beyond the main shell in the X-ray emission, which resembles similar features seen in other SNR (e.g. in Vela SNR), is the interaction of high-speed clumps of SN ejecta with the ambient medium. \citet{Wang02} use two-dimensional hydrodynamic simulations to model the formation of such structures. They find that dense, high-velocity clumps of ejecta may protrude beyond the forward shock as the remnant interacts with the surrounding medium. Shocks moving through the clump at first crush it and then cause it to expand laterally, eventually taking on the shape of a crescent, such as is seen in N63A. The roughly ``triangular hole'' seen in the {\it Chandra} X-ray data near the location of the optical lobes, which is also coincident with the brightest emission in radio, is the result of N63A engulfing a cloud that then absorbs the X-ray emission \citep{Sano19}. 

\subsection{Nature of the Ambient Medium of N63A}
We studied the X-ray spectrum from the outermost shell regions to assign the features of the swept-up ISM (see Figure 2a). The mean values of our fitted model parameters for the outermost shell regions are consistent with \citet{Schenck16} values within statistical uncertainties. Our results show that Ne, Mg, and Si abundances are consistent (within statistical uncertainties) with values of \citet{Russell92}, while O and Fe are lower by a factor of $\sim  2-3$ than \citet{Russell92} values. The median $n_et$ calculated for the swept-up ISM is $0.01\times10^{22}$ cm$^{-2}$.

\subsection{Spatial and Chemical Structure of Ejecta}

The spectral analysis of the observed X-ray spectra for 144 sub-regions in N63A shows O, Ne, and Mg are enhanced in the central parts of the remnant for some directions, while Si and Fe generally shows no enhancement (Figures 7-9). O abundance is enhanced in radial regions A, B, and F at the more than 2 sigma level. However, there is no evidence of an enhanced abundance of O in radial regions C, D, and E. Ne and Mg abundances show no significant enhancement. There is weak evidence for an enhancement of Ne and Mg in the radial regions A and F, while abundances of Ne and Mg are not generally enhanced. There is weak evidence for an enhancement of Ne and Mg in the radial regions A and F, while abundances of Ne and Mg are not enhanced in directions B-E and generally consistent with the mean shell values. Elemental abundances (O, Ne, Mg) for the A and F directions are above the mean shell values up to $r \sim 15''$, while the abundances for the other directions, except the innermost regions, are consistent with the mean shell values. Ejecta gas expanded more in the A and F directions than in the other four directions. Another possibility is that the reverse shock has reached the ejecta sooner in these directions due to the interaction with the molecular cloud.
 
The fact that the element abundance distributions of the spatially selected regions on N63A are not uniform and the elemental abundances in the A and F directions are higher than the measurements in the other directions (B-E) may be caused by an asymmetric explosion as well as by circumstellar medium (CSM). Complex structures in the CSM could have affected the structure of the shocked ejecta. For instance, numerical simulations show that, when the outer blast wave encounters a dense wind shell, a reflected shock is developed \citep[along with the transmitted shock into the shell, e.g.,][]{Dwarkadas05}. Such a shock propagating back to the geometric center of the X-ray emission of the SNR might produce an inwardly increasing density structure. The transmitted shock into the dense shell will enhance the X-ray emissivity of the shocked CSM. The limb brightening in the northeast, northwest and southeast boundaries (see Figure 1), in contrast to the faint and diffuse outer boundary in the southwest, might suggest such an interaction between the blast wave and the dense wind shell. Alternatively, the non-standard structure of the shocked ejecta might have been caused by the inherent complexity in the ejecta. 
For all radial directions, although the $kT$ is slightly higher in the central parts of the remnant, it generally does not show a significant gradient and can be considered constant. The $n_et$ is also generally higher in the inner parts of the remnant and decreases beyond $r\sim 18'' - 20''$. It also does not show a remarkable gradient. The $EM$ exhibits a distribution compatible with the broadband image of the remnant.

\subsection{Inner and Outer Faint Structures on N63A}

The ``crescent-like'' features (Ear 1-3, Tail 1-4) located beyond the main shell in the X-ray emission, which resemble structures seen in the {\it ROSAT} image of the Vela supernova remnant \citep{Aschenbach95}, which have been interpreted as arising from high-speed clumps of SN ejecta interacting with the ambient medium \citep{Warren03}. The origin of ``crescent-like'' structures as ejecta is supported by \citet{Miyata01}, who found an overabundance of Si in their work of Vela shrapnel A. {\it ASCA} and {\it Chandra} data of Vela bullet D show strong O, Ne, and Mg emission \citep{Slane01, Plucinsky01}, suggesting a source as a dense knot of SN ejecta. Nonetheless, \citet{ Plucinsky01} discuss that a nearly solar abundance plasma far from ionization equilibrium is a more likely explanation for their observations. We detect that none of the crescent regions in N63A shows strongly enhanced abundances. Only in the northern crescent region (Ear 1) are the Ne, Mg and Si abundances slightly higher compared to the other sides. The ``crescent-like'' structures of N63A are generally consistent with the median ISM values within the error limits. We can confidently conclude that the ``crescent-like'' structures are not dominated by ejecta. These structures that look softer than the N63A's average in RGB image show the features of the shocked-ISM. Thus, for the high-speed ejecta clump scenario to be correct, the clumps in N63A must be significantly mixed with ISM, perhaps indicating the onset of their disintegration and destruction. 

The element abundances calculated from the spectral analyses of the ``hole-like'' regions in the western inner parts of the remnant (I1-I5) are generally higher than the ``crescent-like'' structures and median ISM values. But it can be considered compatible within the statistical uncertainties. The $n_{\rm e}t$ and $kT$ are higher than the median ISM values. The higher $kT$ and $n_{\rm e}t$ parameters than the ISM values designate that the shock waves generated by the explosion heat and ionize these regions in the inner parts of the remnant. The fact that the element abundances are consistent with the mean shell values also points out that ISM is dominant rather than ejecta for these regions. On the other hand, the $N_{\rm H}$ parameters obtained for these ``hole-like'' regions are approximately 20-40 times higher than the ISM values \citep{Sano19}. These X-ray faint structures in the western inner parts of N63A correspond to active molecular clouds and ionized hydrogen regions in the optical band of the electromagnetic spectrum \citep{Sano19}. The higher $N_{\rm H}$ values calculated for these regions support this finding and also consistent with measurements by \citet{Warren03}.

\subsection{Progenitor Feature and SNR Dynamics}

Based on the mean values of the ejecta elemental abundances measured from individual ejecta regions,  calculated by considering the first six regions in all directions, $r \sim 15''$, we estimate the abundance ratios of O/Ne=$5.7_{-3.6}^{+3.8}$, Ne/Mg=$4.2_{-2.5}^{+3.2}$, O/Mg=$23.0_{-12.7}^{+17.8}$, and O/Si=$26.3_{-13.7}^{+16.1}$. These abundance ratios are in plausible agreement with the core-collapse supernova and hypernova nucleosynthesis models for a $40M_{\odot}$ progenitor with solar or sub-solar (Z = 0.004) metallicity \citep{Nomoto06}. This progenitor mass is consistent with that estimated by \citet{Oey96}.

To estimate the explosion energy and the age of the SNR, we apply self-similar Sedov solutions \citep{Sedov59}. For these purposes, based on the volume emission measure values ($EM = n_{\rm e}n_{\rm H}V$) estimated from the best-fit spectral models of the shell regions we calculate the post-shock electron density ($n_{\rm e}$). For this estimation, we calculated the X-ray emitting volumes ($V$) for each region are listed in Table \ref{tab:dynamic_parameters}. For all shell regions we also assumed a $\sim 0.56$ pc path length (roughly corresponding to the angular thickness of each shell region at 50 kpc) along the line of sight. For a mean charge state with normal composition, we assumed $n_{\rm e}\sim 1.2n_{\rm H}$ (where $n_{\rm H}$ is the H number density) and calculated electron density for all shell region $n_{\rm e} \sim 21.1-36.8 f^{-1/2}{\rm cm}^{-3}$ where $f$ is the volume filling factor of the X-ray emitting gas. The pre-shock H density $n_0$ are listed in Table \ref{tab:dynamic_parameters} assuming a strong adiabatic shock where $n_{\rm H}=4n_0$. Under the assumption of electron-ion temperature equipartition for N63A, the gas temperature is related to the shock velocity ($V_{\rm s}$) as $T=3\hat{m}v_{\rm s}^2/16k$ (where $\hat{m}\sim0.6m_{\rm p}$ and $m_{\rm p}$ is the proton mass). Using electron temperatures, we calculated shock velocity of $V_{\rm s}\sim 652-721$ km s$^{-1}$ and Sedov age of $\tau_{\rm sed}\sim 5,262-5,812$ yr for each shell region (see Table \ref{tab:dynamic_parameters}). The median shock velocity and Sedov age calculated from the six shell region values are $706\pm 25$ km s$^{-1}$ and $5,375\pm 200$ yr, respectively. Our shock-velocity estimate is only a conservative lower limit, therefore our SNR age estimate is an upper limit. Although our age upper limit is not tightly constraining, it is generally consistent with previous estimates of $\sim 4,500$ yr \citep*{Williams06} and 2,000-5,000 yr \citep*{Hughes98, Warren03}. We calculated the corresponding explosion energy of $E_0\sim 6.09-10.43\times 10^{51}$ erg for N63A (see Table \ref{tab:dynamic_parameters}). The median explosion energy is $E_0\sim 8.9\pm 1.6 \times 10^{51}$ erg which is compatible with the explosion energy values given for core-collapse supernovae in the literature \citep{Nomoto06, Vink20}. In addition, the calculated high explosion energy indicates that the remnant may also originate from a hypernova \citep{Janka12}.

\begin{table}
\normalsize
\setlength{\tabcolsep}{4pt}
  \centering
  \caption{Assumed $V$, $n_0$ values for Sedov solutions, and derived dynamic parameters ($V_{\rm s}$, $\tau_{\rm sed}$, and $E_0$) of N63A.}
    \begin{tabular}{cccccc}
\hline
Shell Region& $V$ & $n_0$        & $V_{\rm s}$        & $\tau_{\rm sed}$ & $E_0$\\
& (10$^{55}$ cm$^{3}$) &  (cm$^{-3}$) &  (km s$^{-1})$ & (yrs)  &  ($\times 10^{51}$erg) \\        
\hline
ISM1	& 2.19	& 5.57	& 721 & 5,262 & ~~8.10\\
ISM2	& 2.19	& 6.71	& 715 & 5,306 & ~~9.59\\
ISM3	& 2.19	& 7.67	& 697 & 5,444 & 10.43\\
ISM4	& 1.94	& 6.94	& 652 & 5,812 & ~~8.28\\
ISM5	& 3.54	& 4.40	& 703 & 5,397 & ~~6.09\\
ISM6    & 1.71  & 6.91  & 709 & 5,351 & ~~9.72\\ 
\hline
    \end{tabular}%
    \label{tab:dynamic_parameters}
\end{table}%

\section{Summary \& Conclusion}
We present the results of our extensive analysis of the {\it Chandra} archival data of core-collapse SNR N63A in the LMC. Our detailed spatially-resolved spectral analysis reveals radial profiles of elemental abundances for O, Ne, Mg, Si, and Fe and plasma parameters. We detect an asymmetric structure of central metal-rich ejecta material. The asymmetric distribution of N63A gas is likely caused by an asymmetric explosion of the progenitor, but it should not be ruled out that the ejecta may undergo a non-uniform expansion in interstellar material with different densities. We estimate the explosion energy $E_0\sim 8.9\pm 1.6\times 10^{51}$ erg. This explosion energy estimate is compatible with a Type II SN or hypernova explosion. We also estimate a Sedov age of $\sim 5,400\pm 200$ yr for N63A. 

\section*{Acknowledgements}
We thank the anonymous referee for his/her insightful and constructive suggestions, which significantly improved the paper. This study was funded by Scientific Research Projects Coordination Unit of Istanbul University. Project number: FOA-2018-30716. We would like to thank Dr Jayant Bhalerao for his contributions. This research has made use of data obtained from the Chandra Data Archive and the Chandra Source Catalog, and software provided by the Chandra X-ray Center (CXC) in the application packages CIAO, ChIPS, and Sherpa. This research has made use of NASA's Astrophysics Data System Bibliographic Services. This study is a part of the master's thesis of Emre Karag\"oz.

\section*{Data Availability}
The X-ray data on N63A as described in Section 2 include  Chandra ACIS-S observations and data are available in the Chandra archive (https://asc.harvard.edu/cda/). Processed data products underlying this article will be shared on reasonable request to the authors.





\begin{thebibliography}{99}
\bibitem[Anders \& Grevesse(1989)]{Anders89} 
Anders E., Grevesse N., 1989, Geochimica et Cosmochimica Acta, 53, 197 

\bibitem[Aschenbach, Egger, \& Tr{\"u}mper(1995)]{Aschenbach95}
Aschenbach B., Egger R., Tr{\"u}mper J., 1995, Natur, 373, 587

\bibitem[Badenes et al.(2006)]{Badenes06}
Badenes C., Borkowski K. J., Hughes J. P., Hwang U., Bravo E., 2006, ApJ, 645, 1373 

\bibitem[Bautz et al.(1998)]{Bautz98} 
Bautz M. W., Pivovaroff M., Baganoff F. et al., 1998, SPIE, 3444, 210

\bibitem[Borkowski et al.(2001)]{Borkowski01} 
Borkowski K. J., Lyerly W. J., Reynolds S. P., 2001, ApJ, 548, 820 

\bibitem[Caulet \& Williams(2012)]{Caulet12}
Caulet A., Williams R.~M., 2012, ApJ, 761, 107

\bibitem[Chu \& Kennicutt(1988)]{Chu88}
Chu Y.-H., Kennicutt R.~C., 1988, AJ, 96, 1874

\bibitem[Dickel et al.(1993)]{Dickel93}
Dickel J.~R., Milne D.~K., Junkes N., Klein U., 1993, A\&A, 275, 265

\bibitem[Dwarkadas(2005)]{Dwarkadas05}
Dwarkadas V.~V., 2005, ApJ, 630, 892

\bibitem[Feast(1999)]{Feast99}
Feast M., 1999, IAUS, 190, 542

\bibitem[Foster et al.(2012)]{Foster12}
Foster A. R., Ji L., Smith R. K., Brickhouse N. S., 2012, ApJ, 756, 128

\bibitem[HI4PI Collaboration et al.(2016)]{HI4PI16} 
HI4PI Collaboration, Ben Bekhti N., Floer L., et al., 2016, A\&A, 594, A116

\bibitem[Hughes et al.(1998)]{Hughes98} 
Hughes J. P., Hayashi I., Koyama K., 1998, ApJ, 505, 732

\bibitem[Janka(2012)]{Janka12}
Janka H.-T., 2012, ARNPS, 62, 407

\bibitem[Kennicutt \& Chu(1988)]{Kennicutt88}
Kennicutt R.~C., Chu Y.-H., 1988, AJ, 95, 720

\bibitem[Levenson et al.(1995)]{Levenson95}
Levenson N.~A., Kirshner R.~P., Blair W.~P., Winkler P.~F., 1995, AJ, 110, 739

\bibitem[Lucke \& Hodge(1970)]{Lucke70}
Lucke P.~B., Hodge P.~W., 1970, AJ, 75, 171

\bibitem[Mathewson \& Healey(1964)]{Mathewson64}
Mathewson D.~S., Healey J.~R., 1964, IAUS, 20, 245

\bibitem[Mathewson et al.(1983)]{Mathewson83}
Mathewson D.~S., Ford V.~L., Dopita M.~A., Tuohy I.~R., Long K.~S., Helfand D.~J., 1983, ApJS, 51, 345

\bibitem[McConnachie(2012)]{McConnachie12}
McConnachie A. W., 2012, AJ, 144, 4

\bibitem[Miyata et al.(2001)]{Miyata01}
Miyata E., Tsunemi H., Aschenbach B., Mori K., 2001, ApJL, 559, L45

\bibitem[Nomoto(1982)]{Nomoto82} 
Nomoto K., 1982, ApJ, 253, 798

\bibitem[Nomoto et al.(2006)]{Nomoto06}
Nomoto K., Tominaga N., Umeda H., Kobayashi C., Maeda K., 2006, NuPhA, 777, 424

\bibitem[Oey(1996)]{Oey96}
Oey M.~S., 1996, ApJS, 104, 71

\bibitem[Payne et al.(2008)Payn, White, \& Filipovi{\'c}]{Payne08} 
Payne J.~L., White G.~L., Filipovi{\'c} M.~D., 2008, MNRAS, 383, 1175

\bibitem[Plucinsky et al.(2001)]{Plucinsky01}
Plucinsky P.~P., Smith R.~K., Edgar R.~J., Gaetz T.~J., Slane P.~O., 2001, tysc.conf, 129

\bibitem[Russell \& Dopita(1990)]{Russell90}
Russell S.~C., Dopita M.~A., 1990, ApJS, 74, 93

\bibitem[Russell \& Dopita(1992)]{Russell92}
Russell S. C., Dopita M. A., 1992, ApJ, 384, 508 

\bibitem[Sano et al.(2019)]{Sano19}
Sano H., Matsumura H., Nagaya T., et al., 2019, ApJ, 873, 40

\bibitem[Sato et al.(2007)]{Sato07}
Sato K., Tokoi K., Matsushita K., Ishisaki Y., Yamasaki N.~Y., Ishida M., Ohashi T., 2007, ApJL, 667, L41

\bibitem[Schenck et al.(2016)]{Schenck16}
Schenck A., Park S., Post S., 2016, AJ, 151, 161 

\bibitem[Sedov(1959)]{Sedov59}
Sedov L.~I., 1959, Similarity and Dimensional Methods in Mechanics, New York: Academic Press

\bibitem[Smithsonian Astrophysical Observatory(2000)]{ds9cite}
Smithsonian Astrophysical Observatory. 2000, SAOImageDS9: A utility for displaying astronomical images in theX11 window environment.  http://ascl.net/0003.002

\bibitem[Shull(1983)]{Shull83}
Shull P., 1983, ApJ, 275, 592

\bibitem[Slane et al.(2001)]{Slane01}
Slane P., Hughes J.~P., Edgar R.~J., Plucinsky P.~P., Miyata E., Tsunemi H., Aschenbach B., 2001, ApJ, 548, 814

\bibitem[Tsujimoto et al.(1995)]{Tsujimoto95}
Tsujimoto T., Nomoto K., Yoshii Y., Hashimoto M., Yanagida S., Thielemann F.-K., 1995, MNRAS, 277, 945

\bibitem[van den Bergh \& Dufour(1980)]{vandenBergh80}
van den Bergh S., Dufour R.~J., 1980, PASP, 92, 32

\bibitem[Vink(2020)]{Vink20}
Vink J., 2020, Physics and Evolution of Supernova Remnants, Springer Nature Switzerland AG, ISBN: 978-3-030-55231-2

\bibitem[Wang \& Chevalier(2002)]{Wang02}
Wang C.-Y., Chevalier R.~A., 2002, ApJ, 574, 155

\bibitem[Warren et al.(2003)Warren, Hughes, \& Slane]{Warren03}
Warren J.~S., Hughes J.~P., Slane P.~O., 2003, ApJ, 583, 260

\bibitem[Westerlund \& Mathewson(1966)]{Westerlund66}
Westerlund B.~E., Mathewson D.~S., 1966, MNRAS, 131, 371

\bibitem[Williams et al.(2006)Williams, Chu, \& Gruendl]{Williams06}
Williams R.~M., Chu Y.-H., Gruendl R., 2006, AJ, 132, 1877

\bibitem[Yamaguchi et al.(2014)]{Yamaguchi14}
Yamaguchi H., Badenes C., Petre R., et al., 2014, ApJL, 785, L27


\end{thebibliography}

\bsp	
\label{lastpage}
\end{document}